\documentclass[12pt]{article}
\usepackage{graphicx, epsfig}
\usepackage{amsmath}
\usepackage{latexsym}
\usepackage{amstext}
\usepackage{array}
\usepackage{multirow}
\usepackage{tikz}
\usetikzlibrary{arrows}
\usepackage{changepage}
\usepackage{xfrac}
\usepackage{subcaption}
\usepackage{hyperref}

\newcommand{\ds}{\displaystyle}
\newcommand{\beq}{\begin{eqnarray}}
\newcommand{\eeq}{\end{eqnarray}}
\newcommand{\beqq}{\begin{eqnarray*}}
\newcommand{\eeqq}{\end{eqnarray*}}

\begin{document}
\begin{center}
\Large \textbf{Segmentation algorithms and modeling of recurrent bursting events in neuronal and glial time series}\\
\vspace{0.25cm}
{\bf Springer Nature book about Neuromethods}\\
\vspace{0.25cm}
\normalsize
L. Zonca\footnote{Group of Applied Mathematics and Computational Biology, Ecole Normale Sup\'erieure, PSL University, Paris, 46 rue d'Ulm, 75005 Paris, France. New address: Center for Brain and Cognition, University Pompeu Fabra, Barcelona}, E. Dossi\footnote{Neuroglial Interactions in Cerebral Physiopathology, Center for Interdisciplinary Research in Biology, Coll\`ege de France, CNR UMR 7241, INSERM U1050, Labex Memolife, PSL Research University, Paris, France}, N.Rouach$^{2}$  and D. Holcman$^{1,}$\footnote{Churchill College, Cambridge University, CB30DS UK. email: david.holcman@ens.fr}
\end{center}
\begin{abstract}
Long-time series of neuronal recordings are resulting from the activity of connected neuronal networks. Yet how neuronal properties can be extracted remains empirical. We review here the data analysis based on network models to recover physiological parameters from electrophysiological and calcium recordings in neurons and astrocytes. After, we present the recording techniques and activation events, such as burst and interburst and Up and Down states. We then describe time-serie segmentation methods developed to detect and to segment these events. To interpret the statistics extracted from time series, we present computational models of neuronal populations based on synaptic short-term plasticity and After hyperpolarization.  We discuss how these models are calibrated so that they can reproduce the statistics observed in the experimental time series. They serve to extract specific parameters by comparing numerical and experimental statistical moment or entire distributions. Finally, we discus cases where calibrated models are used to predict the selective impact of some parameters on the circuit behavior,  properties that would otherwise be difficult to dissect experimentally.
\end{abstract}

{\bf Keywords:} Signal segmentation; Algorithms; Computational models; Deconvolution; Nanopipette; Bursting; Correlation; AfterHyperPolarization; Synaptic short-term plasticity; Neuronal network; Parameter estimation; Optimization.

\section*{Introduction}
Neuronal recordings reveal complex network activation patterns. Although these patterns  are clearly appearing in the data, we still need to segment them. This is the preliminary step for analyzing, decoding and thus better clarifying how neuronal networks impact the brain function \cite{kandel2000}. Electrophysiological recordings with patch-clamp \cite{hille1978} provide a direct access to the voltage dynamics of the soma or dendrites of a single neuron (Fig. \ref{MEA-electrophy}). On one hand, this technique revealed transient behaviors, such as spiking \cite{stuart1997action,markram1997regulation}, persistent fluctuations of the membrane potential called Up and Down states \cite{konnerth2006} or bursting \cite{steriade2006burstingReview}. On the other hand, indirect recordings of neuronal activity such as local field potentials or multiple electrode array (MEA), are used to record population ensembles containing tens to millions of neurons and thus providing access to global activity (Fig. \ref{MEA-electrophy}). \\
Another method of probing network activity consist in recording transient calcium events based on recording fluorescent dyes. Although with a much coarser time resolution, calcium recordings reveal spatial synchrony between neurons \cite{kenet2003spontaneously,Cossart2003}. The neuronal code embedded into the signal from single recordings contains several components, mixing intrinsic background instrinsic fluctuations and instrumental noise. In that context, spontaneous events, resulting from the background fluctuations, can be difficult to differentiate from evoked events \cite{chen2013reactivation}. To overcome theses difficulties, methods are combining signal processing to extract statistics from the data with the help of mathematical modeling to interpret data. We propose to review here some of these approaches.\\
The present review is organized as follows: first, we describe briefly the electrophysiological and calcium fluorescence recording techniques, as well as the signals they provide. Second, we present time segmentation procedures to distinguish large amplitude events, characterized by sharp transitions, from the background noise. We recall how to construct and interpret the statistics obtained from the segmentation, with a focus on the bursting events but also on the inter-events intervals. The present approach is generic as it is used to analyze spontaneous or evoked activity. Ongoing cortical neuronal activity in the absence of sensory inputs or with sensory stimulation often reveal similar distributions \cite{chen2013reactivation}. Finally, in the third part, we review parsimonious models based on short-term synpatic plasticity, offering a reduced number of variables. These models are used to interpret the distributions of burst and interburst intervals and Up and Down states. Once calibrated, these models allow to evaluate role of specific variables, which could be otherwise difficult to disentangle experimentally.
\section{Electrophysiological and calcium time series recordings}
\subsection{Patch-clamp, local-field potential and MEA}
\begin{figure}[http!]
\centering
\includegraphics[scale=0.8]{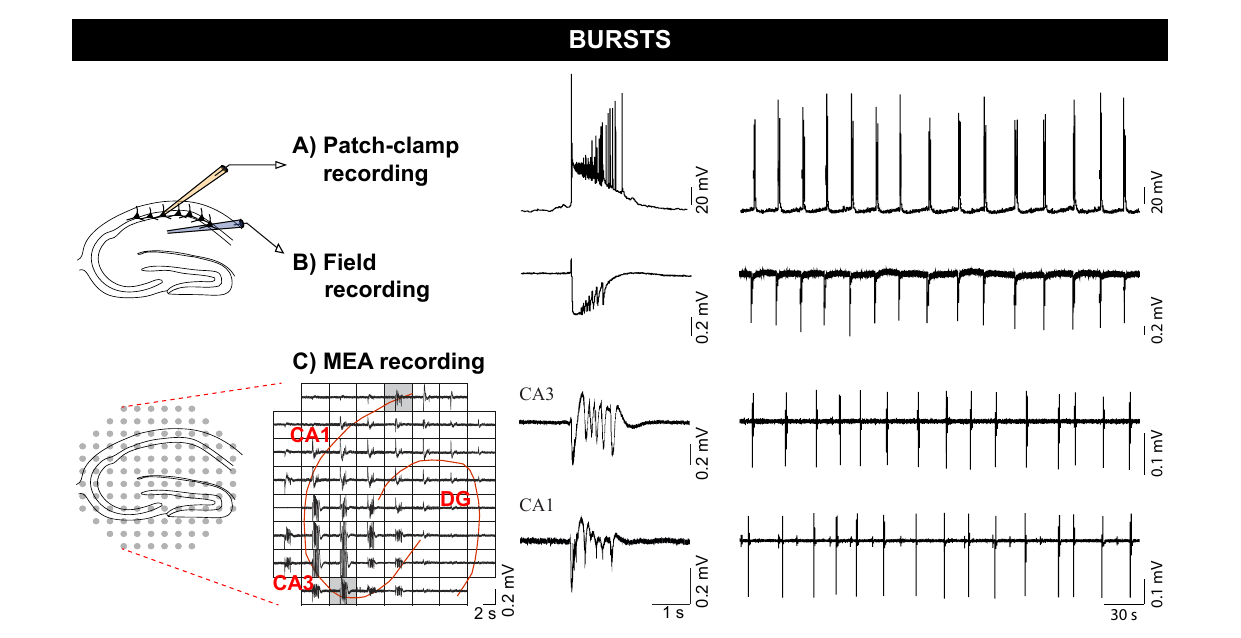}
\caption{\textbf{Bursting activity recorded in mouse hippocampal slices with different electrophysiology techniques.} {\bf A.} Patch-clamp recording for a pipette penetrating the cell {\bf B.} local field potential (LFP) when the tip of the pipette is immersed in the bath. {\bf C.} Multi-electrode array (MEA) recordings consisting of multiple electrodes recording the sum for the activity of small and large neuronal networks. This recording allows to follow the center of an epiletpic focus. Left, schematic representations of patch-clamp and field pipette positions for recording and MEA array. Middle, enlarged view of single bursts recorded with the three techniques. Right, representative continuous recordings of recurrent bursts (5-minutes time windows). (Modified from \cite{Rouach_CxKO}, and unpublished data).} \label{MEA-electrophy}
\end{figure}

The activity of the brain spans over multiple temporal and spatial scales, that require an adequate set of technologies to be addressed. In the past decade, robust recording approaches have revealed transient events in individual neurons and their organized network. Electrophysiology recordings capture a wide spectrum of neuronal phenomena, from single neuron spiking to slow and fast network oscillations. Different electrophysiological techniques can be used to study brain neuronal activity, among which patch-clamp, local field potential (LFP) or multi-electrode array (MEA) recordings.\\
The patch-clamp technique is based on the use of a glass micropipette to form high resistance seals on tiny patches of cell membrane \cite{kornreich2007patch}, allowing high-resolution recording of currents from single ionic channels, or from the entire population of ionic channels, on a cell membrane \cite{sachs19839}. This powerful technique thus provides intracellular recordings of neuronal activity at the single cell level, but its use is limited to a few neurons per experiments. During patch-clamp, an electrode is inserted in the soma of a chosen neuron, and is now routinely used to obtain long-time stable recordings for hours. Examples of double recordings in the hippocampus show a first patch electrode inside a neuron (Fig. \ref{MEA-electrophy}A) and a field recording, called local field potential (LFP), where the pipette is simply immersed in the tissue (Fig. \ref{MEA-electrophy}B). We recall that LFP recording is an electrophysiology technique that focuses on the activity of large populations of cells. Indeed, LFP data mirror the spatially weighted electrical activity in the proximity of the recording electrode \cite{mazzoni2013information,pesaran2009uncovering}, including excitatory and inhibitory post-synaptic potentials, non-synaptic calcium spikes as well as afterpotential of somatodendritic spikes and voltage-gated membrane oscillations \cite{oren2010currents,obien2015revealing,buzsaki_brain_2012}. Thus, LFP measurements detect a variety of subthreshold activity close to the electrode and provide a view of local extracellular network activity at a single site \cite{mazzoni2013information}. From long-time recordings, it is possible to sample short bursting events of few seconds, with different electric signatures depending on the recording techniques, but with similar time scale.\\
The key advantage of multi-electrode array (MEA) recording compared to techniques mentioned above is the capability to record (and stimulate) neuronal activity (extracellular local field potentials and extracellular action potentials) at multiple sites simultaneously \cite{obien2015revealing}, thus allowing the study of neuronal network activity pattern and propagation. Microelectrodes embedded in large arrays detect changes in the extracellular field caused by ionic flows through the closest neurons and nearby cells \cite{buzsaki_brain_2012,anastassiou2013biophysics}, and the amplitude and shape of detected signals depend on the orientation and distance between the neuronal sources and the recording electrodes \cite{nunez2006electric,obien2015revealing} (Fig. \ref{MEA-electrophy}C). Based on the spatial precision gained by the MEA, events with decaying amplitudes can be tracked, enabling to find regions where a burst can spread and thus and to study for example epileptic focus. MEAs have a similar time-resolution as patch-clamp and LFP but the burst dynamics results from  the summation of many groups of neurons. Other improved recording technics include tetrodes to recored extracellular field potentials invivo, multiple electrodes, such as the neuropixels probe features using 960 recording electrodes, allowing to recored more that 700 neurons simultaneously from  brain regions in an awake mouse. Similarly, neuroSeeker probes can also record from 1000 of electrodes in vivo \cite{hong2019novel}.
To conclude transient patterns such as bursting can vary depending on the recording methods, but not the associated time scale, allowing to define a precise time interval and thus to recover duration statistics.
\subsection{Bursting events revealed by calcium dyes}
\begin{figure}[http!]
\centering
\includegraphics[scale=0.8]{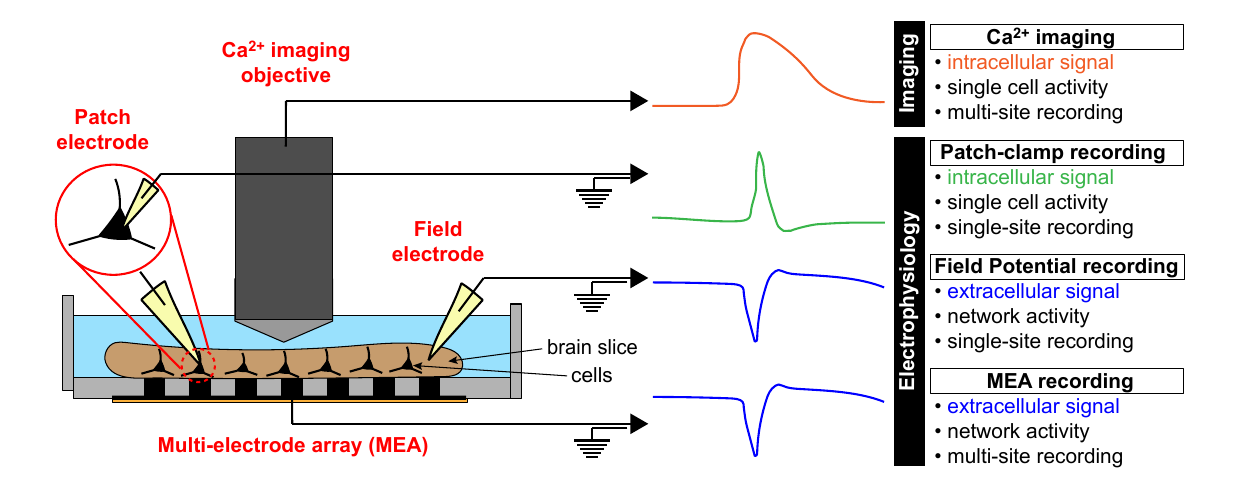}
\caption{\textbf{Comparison between voltage and dyes recording techniques.} Patch-clamp, field potential and MEA recordings versus the slow Ca2+ imaging event.} \label{Comparing}
\end{figure}

Contrary to the fast time-resolved electrophysiological recordings presented in Fig. \ref{MEA-electrophy}, bursting events are also revealed by slower calcium transient events. This approach is well discussed in \cite{savtchenko2018disentangling} for calcium imaging and also most recently for voltage dyes \cite{holcman2015new,zhang2016optical,liu2020recent}. The time courses $f_{dyes}(t)$ of fast events results from the convolution between the dyes response time scale $g_{intrin}(t)$ and the signal itself $C(t)$, which could last much longer than the event. This slower response leads to possible temporal imprecision in the beginning and the end of a burst or a spike, as shown in Fig. \ref{Comparing}. The general problem is often to recover the concentration or voltage from the slow dyes signal. This question is well known in signal processing and is equivalent to finding the kernel $C$ of the following equation
\beq
f_{dyes}(t) = \int_{0}^{t} C(s)g_{intrin}(t-s)ds +N(t),
\eeq
where $N(t)$ is a noise source. A solution based on fitting the empirical curve by an analytical function was proposed in \cite{cartailler2018deconvolution}. This allows  to recover a fast voltage from slower dye responses.
\section{Segmentating, correlation and causal analysis of long time-series:} 
Segmenting transient bursting events automatically in electrophysiological recordings is the first step for collecting large amount of samples and for building the assocaited statistics. Segmentation methods \cite{cohen2014analyzing} consist in isolating transient events of interest such as spikes, bursts or interburst intervals. This segmentation is possible due to the structure of a burst made of sharp transients with a significant in amplitude change compared to the background signal. Segmenting events with an amplitude similar to the background signal would be quite difficult, unless a difference can be identified in the frequency domain, which is the case in ripples \cite{buzsaki_brain_2012}. Several examples of fast transients with large amplitudes that differentiate bursting from the background signal and also hyperpolarization from resting phases are shown in Fig. \ref{MEA-electrophy}. We shall now describe the successive steps to segment bursting events in patch-clamp and MEA time series.
\subsection{Burst detection and segmentation in patch-clamp recordings}\label{ss:segmentation}
The segmentation procedure is to detect burst, AHP and interburst intervals in neuronal patch-clamp recordings.
\begin{figure}[http!]
\centering
 \includegraphics[scale=0.8]{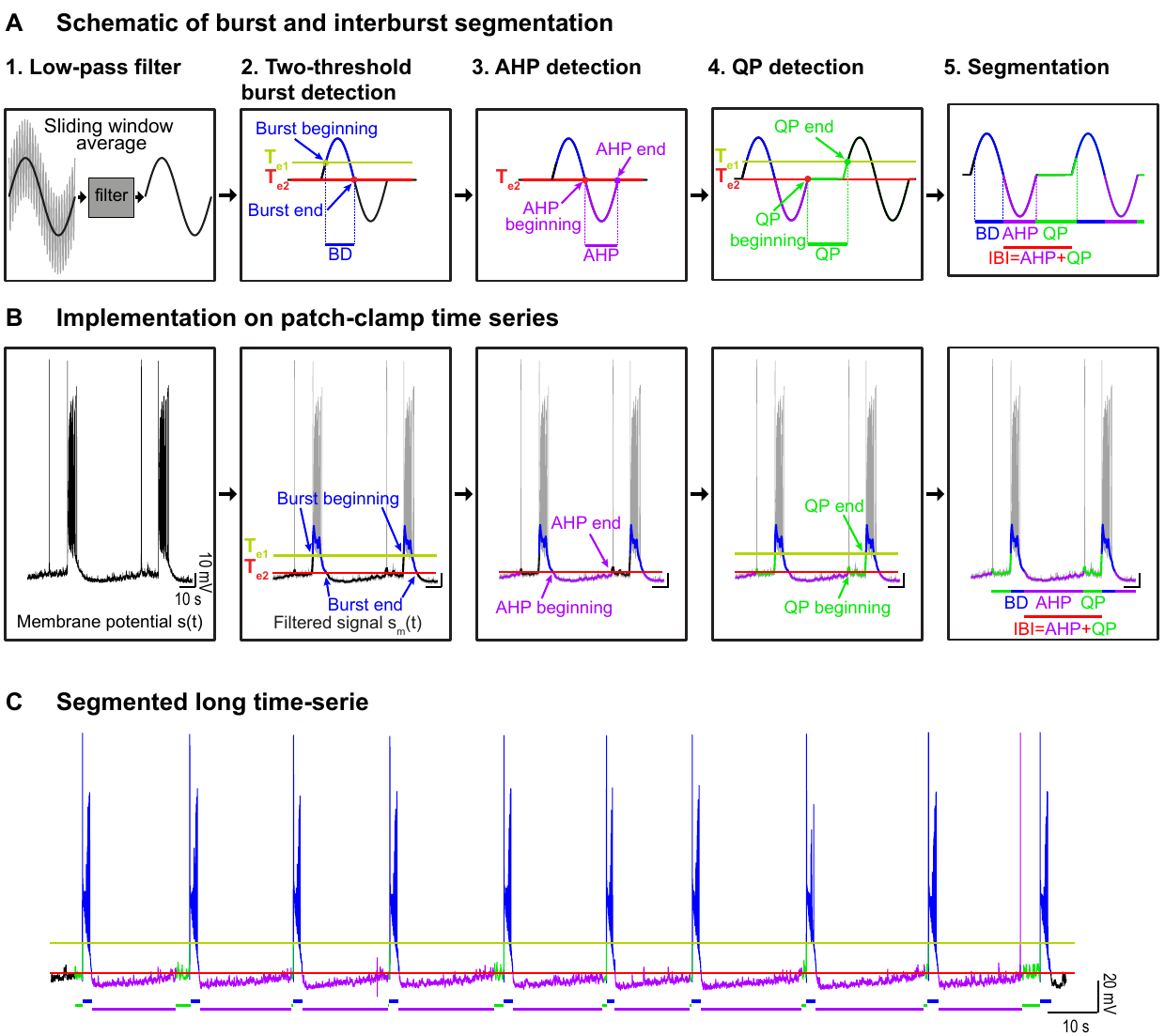}
\caption{\textbf{Burst detection principle.} \textbf{A. Burst and interburst segmentation algorithm.} 1. The membrane potential signal is low-pass filtered using a sliding window of length $T_w = 1s$. 2. A burst is detected using the two thresholds $T_{e,1}$ (beginning) and $T_{e,2}$ (end) on the filtered signal $s_m$. 3. AHP begins at the end of the burst and lasts until $s_m$ is back above resting membrane potential ($=T_{e,2}$) 4. QP begins at the end of AHP until the next burst. 5. Extraction of burst durations, (BD,blue), AHP duration (magenta) and quiescent phases (QP, green). IBI (red) is composed of AHP and QP. \textbf{B. Implementation on patch-clamp time-series.} Step by step application of the algorithm described in A on electrophysiological data. \textbf{C.} $200$s electrophysiological recording segmented into burst (blue) AHP (magenta) and QP (green) periods using the algorithm described in A.} \label{burstDetect}
\end{figure}
\subsubsection{Initial Low-pass filtering}
The first step of burst segmentation consists in removing from the membrane potential $s(t)$ the high frequency components often generated by fast opening and closing of voltage activated channels \cite{hille1978,gribkoff2008structure}. This step is achieved by applying a low-pass filter to $s(t)$, using a sliding window average: for each time point $t$, the filtered signal $s_m(t)$ (fig. \ref{burstDetect}A1) is obtained by taking the average of the unfiltered signal $s$ over a window of size $T_w=1s$ centered at time $t$:
\beq
s_m(t)= \cfrac{1}{T_w}\sum_{i=t-T_w/2}^{t+T_w/2}s(i).
\eeq
This procedure is easy to use, fast at low computational cost. Other averaging methods include Savitzky–Golay or kernel filters. In addition, increasing the filter frequency is simply obtained by increasing the size of the window. This is an advantage compared to classical low pass filters such as Butterworth versus Chebichev, Hermite \cite{jaffard_wavelets_nodate,OPT}, where the order of the filter has to be specified, as well as a cutting frequency. These filters also introduce some distorsions in the signal.
\subsubsection{Detecting the burst initiation}
The burst initiation is detected by finding the first time $\tau^i$ when the the filtered signal $s_m$ reaches a threshold $T_{e1}$ (Fig. \ref{dataAHP}A2, light green), i.e.
\beq
s_m(\tau^i)=T_{e1}.
\eeq
The threshold $T_{e1}$ is chosen relatively to the amplitude of the signal variations as
\beq
T_{e1}= \cfrac{T_{e2} + \max_{t\in [0,T]}s_m(t)}{2},
\eeq
which uses the average between the maximum value of the filtered signal and the resting membrane potential $T_{e2}$ (red line). In general, there is no automatic procedure to detect the membrane resting potential, which can vary over different traces.\\
When the voltage contains mostly bursting and AHP events, with a baseline in the range $R_{mp}=[-55 -65]$mV, one possibility is to impose that the threshold $T_{e2}$ is obtained by averaging the signal inside the range $R_{mp}$ that is
\beq
T_{e2}=\frac{\int_{s_m \in [-55 -65] }s_m(s)ds}{\int_{s_m \in [-55 -65] }ds}.
\eeq
The burst initiation occurs shortly before the detection time $\tau^i$, where the signal $s_m$ has reached the threshold $T_{e1}$, when $s_m$ was at resting state. However, due to the stiff slope, we can neglect this delay of the order of a few milliseconds in comparison with the duration of the burst, typically of the order of hundreds of ms to seconds.
\subsubsection{Detecting the burst termination and AHP phase}
\begin{figure}[http!]
\centering
\includegraphics[scale=0.8]{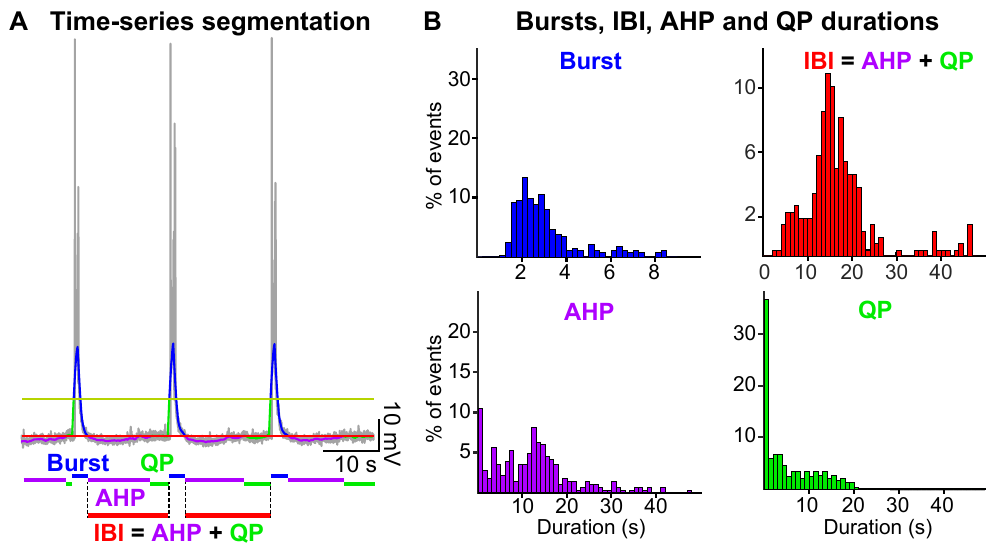}
\caption{\textbf{A.} Segmentation of the traces into burst (blue), AHP (pink) and QP (green), scalebar: 10 s, 10 mV. \textbf{B.} Distributions of bursts (blue), IBI$=$AHP$+$QP (red), AHP (pink) and QP (green) durations.}\label{dataAHP}
\end{figure}

To determine the burst end, we consider the first time when the signal $s_m$ decreases to its resting value defined by $s_m(\tau^e)=T_{e2}$ (Fig. \ref{burstDetect}A2, red line). The end time point $\tau^e$ of the burst also corresponds to the beginning of the AHP phase that lasts until the signal $s_m$ has increased back to its resting value $T_{e2}$ at a further time $\tau^a$, such that $s_m(\tau^a)=T_{e2}$, with the condition that  $\tau^a> \tau^e$ (Fig. \ref{burstDetect}A3, magenta trace). Finally, this time segmentation allows to defin a quiescence phase QP, between the end of the AHP at time $\tau^a$ and the initiation of the next burst (Fig. \ref{burstDetect}A4, green trace). In summary, the $n^{th}$ burst duration is the difference between the termination and initiation times
\beq
BD_n=\tau_n^e-\tau_n^i.
\eeq
The collection of durations $(BD_1,..BD_n)$ (blue), the AHP durations $AHPD_n=\tau_n^a-\tau_n^e$, (magenta) and the quiescent phases $QP_n=\tau_{n+1}^i-\tau_n^a$ (green) define the empirical distributions of segmented traces (Figs. \ref{burstDetect}C and \ref{dataAHP}A). These distributions are obtained from segmentating path-clamp recordings, as shown in Fig. \ref{dataAHP}B.
\subsection{Segmenting bursting in MEA recordings}
The method described above can also be applied with little modifications to MEA recordings. However, the AHP cannot be identified in these recordings, allowing to simplify the segmentation procedure. We explain briefly how the steps have to be adjusted: The starting point is to apply a low-pass filter to the input field recordings using a sliding time window with a length of $T_w=0.4s$ leading to an ouput signal $s_{m,MEA}$. The bursts and interburst intervals are then detected using the absolute value of the filtered signal $|s_{m,MEA}|$. Since the MEA baseline is zero, the burst initiation time $\tau^i$ is obtained when the signal $|s_{m,MEA}|$ reaches one third of its maximum value
\beq
|s_{m,MEA}|(\tau^i)=\cfrac{\max_{[0,T]}(|s_{m,MEA}|)}{3}.
\eeq
Similarly, the end time $\tau^e$ of the burst is found under the condition
\beq
|s_{m,MEA}|(\tau^e)=T_{end}=\cfrac{max(|s_{m,MEA}|)}{15}.
\eeq
This empirical threshold $T_{end}$ allows to identify the end of the burst. It corresponds to the time at which the signal enters into a fluctuation regime around zero, where the noise is dominant in the interburst interval. In general, it would be possible to define this threshold by considering the standard deviation of the signal in between bursts:
\beq
\sigma_{IBI} =  \frac{\int_{s_{m,MEA}(s) \in IBI_i}(s_{m,MEA}(s)-\bar{s_{m,MEA}})^2ds}{\int_{s_{m,MEA}(s) \in IBI_i}ds}
\eeq
where $IBI_i$ are the interbust intervals. This standard deviation can be computed by a simple thresholding with one threshold to detect the burst events. The complementary of these events define the signal where the standard deviation should be computed. In practice, we can replace $\cfrac{max(|s_{m,MEA}|)}{15}$ by $\sigma_{IBI}$. {The Matlab code for the above segmentation algorithms is available for download at \href{www.bionewmetrics.org}{\url{bionewmetrics.segmentation_tools}}. Each algorithm is provided as a Matlab function that takes the data time-series as an input and where the threshold parameters and filtering can be adjusted.}
\subsection{Segmenting intracellular recordings into Up and Down states}
\begin{figure}[http!]
\centering
\includegraphics[scale=0.8]{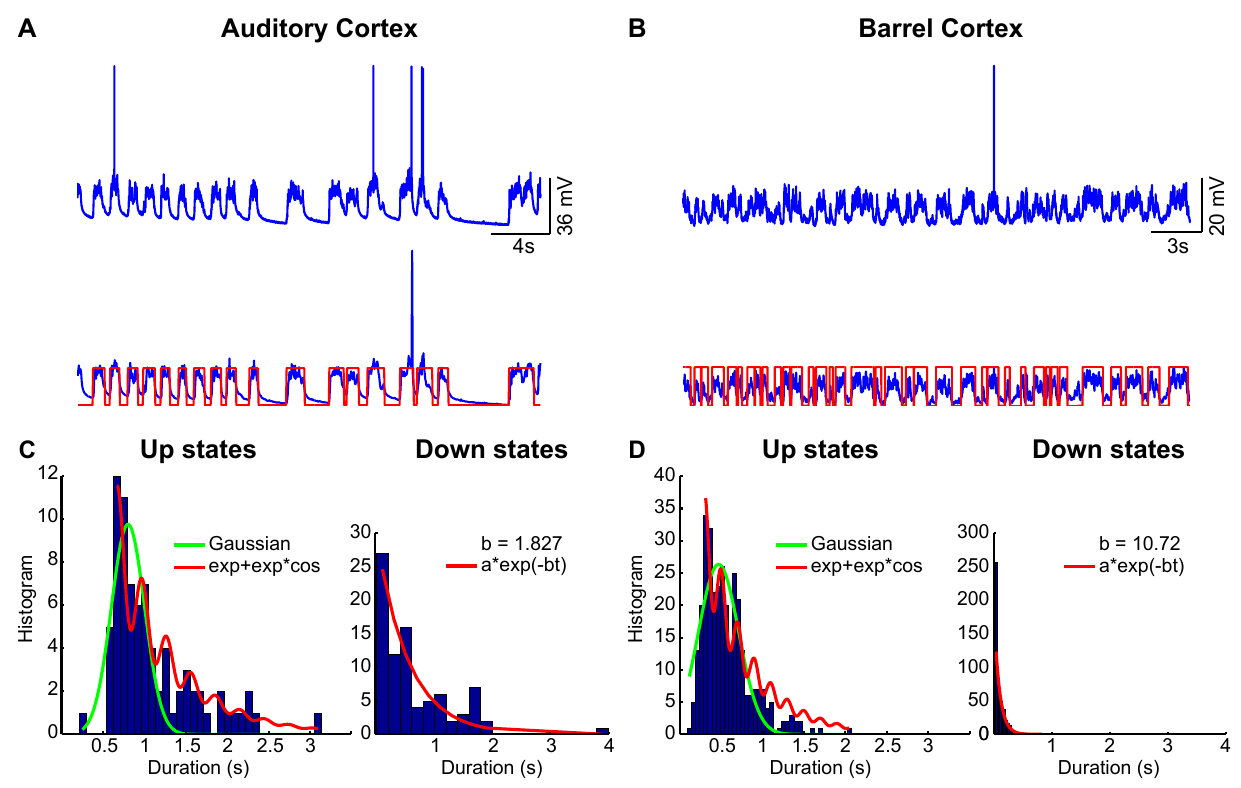}
\caption{{\bf A-B.} Segmentation between Up and Down states (red) of intracellular recordings (blue) in the auditory (A) and barrel (B) cortex, data from \cite{chen2013reactivation}. {\bf C-D.} Distributions of the Up (left) and Down (right) states durations fitted with an analytical curve for both datasets.} \label{Updownsegmentation}
\end{figure}

We present here another application of the segmentation method asociated with the changes of the membrane potential of connected neurons, that can oscillate between two domainant values called Up (depolarized) and a Down (hyperpolarized) states. These two-state oscialltion is due to intrinsic channel properties combined with synaptic dynamics \cite{destexhe2007corticothalamic}. In the past 20 years, neuronal information processing in the cortex has been associated with these Up and Down state dynamics \cite{Cossart2003,chen2013reactivation}: Up-state is often associated with a direct processing of an external stimulation response such as auditory, visual or somatosensory (wisker deviation), but could also appear spontaneously \cite{kenet2003spontaneously} in the form of an organized activity \cite{yang2018simultaneous}. Up and Down states are recognized by the two peaks in the voltage distribution. Interestingly, segmenting intracellular recordings into periods of Up and Down states falls into the method described above. This segmentation allows to build the statistics for their durations, as shown for the auditory and barrel cortex (Fig. \ref{Updownsegmentation}A-B). Interestingly, the segmentation of the intracellular recordings for various neuronal types revealed that the distributions of time in the Up state shows multiple peaks \cite{holcman2014oscillatory,dao2015synaptic}, a sign of a non-Poissonian distribution (Fig. \ref{Updownsegmentation}C-D left). This is however not the case for the distribution of times in the Down-state (right). Extracting from the statistics of these distribution relevant paramter for neuronal activativy will be discuss in section \ref{s:modeling}.
To interpret the fast oscillations in the Up state, a modeling approach is needed as we shall see in Section \ref{s:modeling}.
\begin{table}
\begin{center}
\begin{tabular}{|r|c|c|c}
\hline
& Patch-clamp & MEA \\ \hline
Low-pass filter window $T_w$ & $1$s & $0.4$s \\
Burst detection threshold $T_{e1}$ & $\cfrac{T_{e2} + \max_{t\in [0,T]}s_m(t)}{2}$ & $\cfrac{\max_{[0,T]}(|s_{m,MEA}|)}{3}$ \\
End of burst \& AHP threshold $T_{e2}$ & $\cfrac{\int_{s_m \in [-55 -65] }s_m(s)ds}{\int_{s_m \in [-55 -65] }ds}$ & $\cfrac{max(|s_{m,MEA}|)}{15}$\\
\hline
\end{tabular}
\end{center}
\label{tableSeg}
\caption{Experimental time-series segmentation parameters.}
\end{table}
\subsection{Time correlation between induced and spontaneous burst and interburst events}
\begin{figure}[http!]
\centering
\includegraphics[scale=0.75]{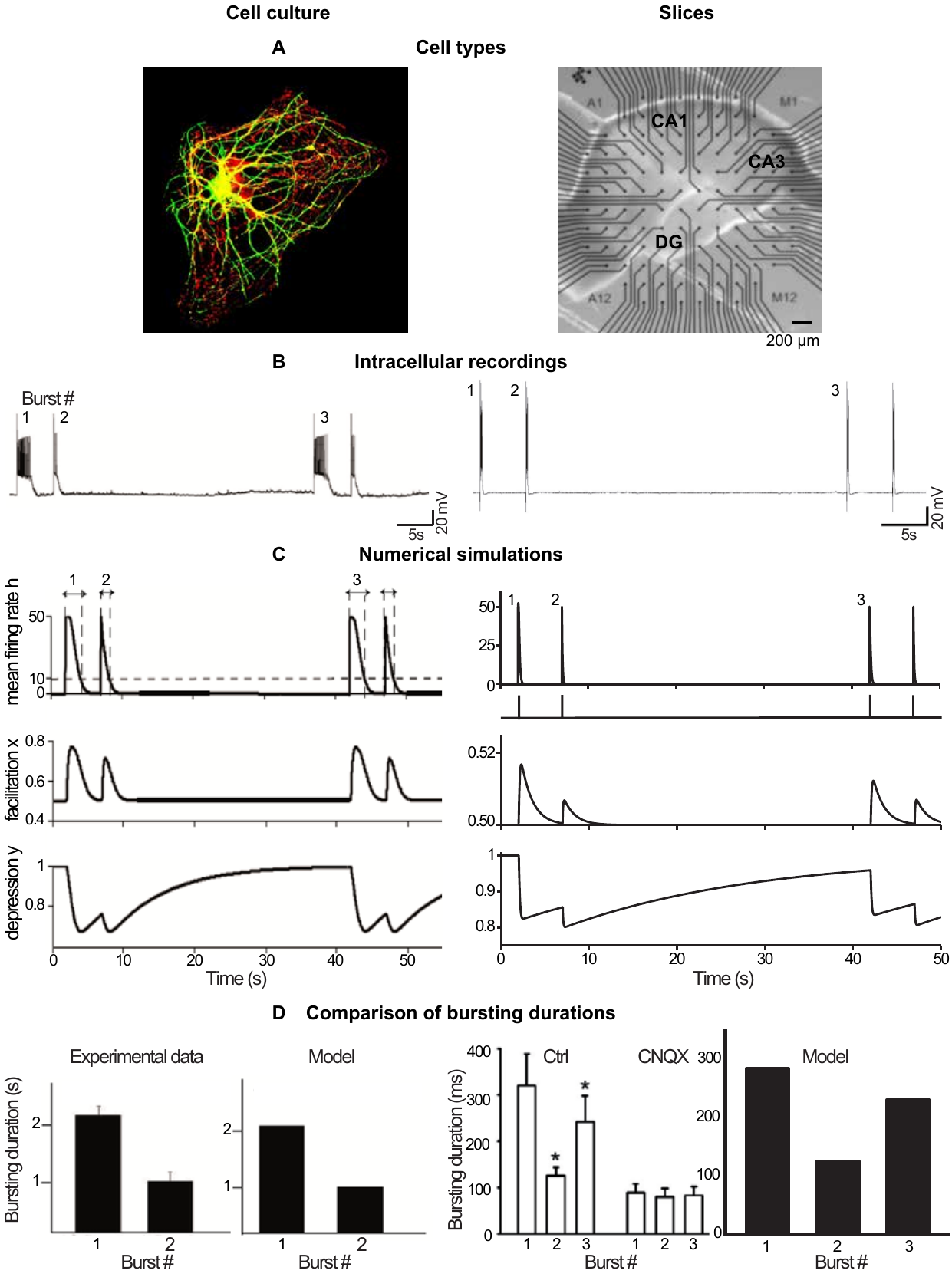}
\caption{{\bf Bursting reverberation times in neuronal cultures and slices.} {\bf A.} Cultured neurons (left) and mice hippocampal slices (right). {\bf B.} Intracellular recordings with evoked bursts 5 s appart (bursts 1 and 2) and then 30 s and 35 s later (bursts 3 and 4) in both cell types (cultures and slice). {\bf C.} Simulations of model \eqref{sysDaop} reproducing the successive bursts calibrated on both datasets, mean firing rate h (upper), facilitation x (center) and depression y (lower). {\bf D.} Comparison of the successive burst durations in data and model experiments.}
\label{reverberation}
\end{figure}

In this section, we discuss how burst segmentation is used to study the correlation between successive events. We start with burst reverberation \cite{cohen2011network,DaoDuc2015}, which consists in the activation of a single neuron that can trigger the firing of an entire well connected network for seconds due to synaptic facilitation and depression dynamics (Fig. \ref{reverberation}A-B). This effect was first reported in cultured islands \cite{Cohen_Segal2009} and then found in slices. Interestingly, the bursting reverberation revealed a long depression time scale that affects the second burst triggered a few seconds later following the first one, leading to a short-term memory effect. This effect was also found in slices, but with a different time scale (hundreds of ms instead of seconds, Fig. \ref{reverberation}C-D), possibly suggesting that the number of neurons involved in the reverberation could play a role in defining the time scale. We will further discuss the interpretation of this phenomenon using a modeling approach in section \ref{s:modeling}.
In induced bursts, the strong correlation between the durations of the second burst which is always shorter than the first one is due to the synaptic dynamics, as removing temporarily the synaptic transmission abolished all bursts \cite{cohen2011network}. In addition, the causal relation between the bursts is well accounted for by a mean-field model based on synaptic depression-facilitation\cite{DaoDuc2015} (see eq. \ref{sysDaop} in section \ref{s:modeling}) as shown in Fig. \ref{reverberation}. This correlation between consecutive bursts demonstrates the role of long lasting synaptic depression. Interestingly, increasing the network connectivity from few neurons in an island to well connected dense networks slices is associated with a reduction of the correlation time \cite{DaoDuc2015}, probably due to a stronger network activity, that could replete faster the available vesicles.\\
\begin{figure}[http!]
\centering
\includegraphics[scale=0.8]{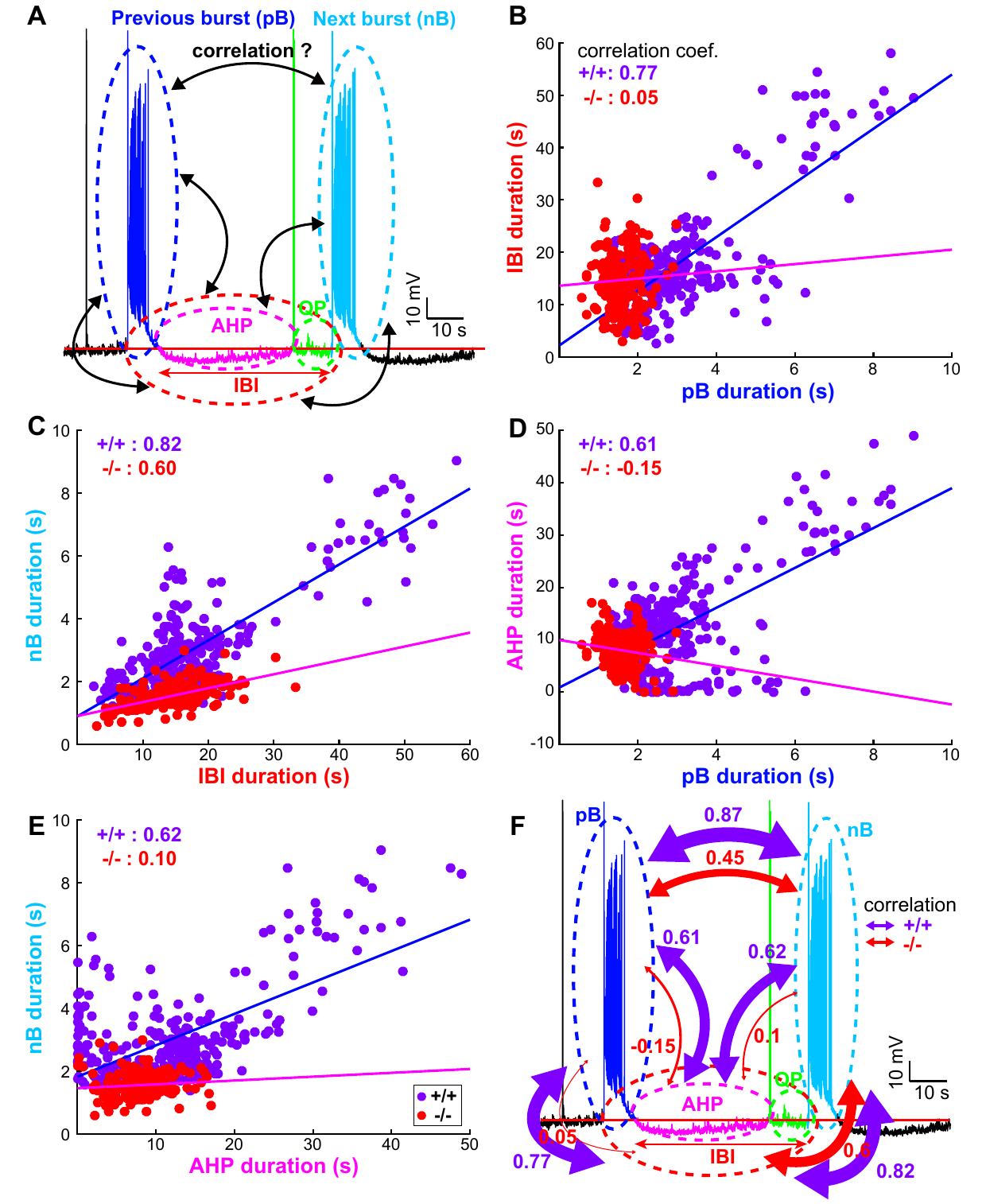}
\caption{{\bf Correlations between and across burst and interburst intervals} {\bf A.} Segmented time-series showing the time epochs used for the correlation study. {\bf B-E.} Correlation plots for Burst occuring in the WT (+/+, purple) and in Connexin KO (-/-, red) between Interburst Interval (IBI) vs previous Burst (pB), next Burst (nB) vs IBI, AHP vs pB and nB vs AHP. {\bf F.} Correlation coefficients (arrows) between the different epochs for (+/+, purple) and (-/-, red), the arrow width represents the correlation strength.}\label{corrFig}
\end{figure}
In parallel to bursting, induced by an external stimulation, it is also relevant to study the spontaneous bursts generated in a network and collect the statistics. For that goal, the bursting events should be collected by segmenting electrophysiological time series as described in section \ref{ss:segmentation} (Fig. \ref{corrFig}A). To increase our understanding of bursting, a genetic perturbation has been applied in the network \cite{chever2016astroglial} to remove the connexin-30 and 43, the role of which is to connect astrocyte into a network and to allow passive diffusion through gap junctions. This modification is used to study how altering the astrocytic network can change the dynamics of spontaneous bursting dynamics occurring in neuronal networks. In practice,  we shall compare the wild type (WT) dataset, with no genetic modification with the one from the KO.\\
To collect a large number of bursting, AHP and IBI events, we apply the segmentation introduced in subsection \ref{ss:segmentation} so that to conduct a correlation analysis. The possible correlations are for the burst durations with the preceding or the next interburst interval and also burst duration vs the previous or the next burst.\\
A Pearson correlation analysis between most features in the KO electrophysiology traces compared to the WT case (Fig. \ref{corrFig}B) reveals specific trends: a high correlation between AHP duration and the preceding (resp. following) burst duration in the WT (Fig. \ref{corrFig}E correlation coefficient: 0.615, resp. F correlation coefficient: 0.619) which is significantly reduced in the KO case (Fig. \ref{corrFig}E correlation coefficient: -0.154, resp. F correlation coefficient: 0.099). Regarding the total interburst duration vs burst durations, the correlation is also reduced in the KO (preceding burst: fig. \ref{corrFig}C correlation coefficient: 0.766 for the WT and 0.052 for the KO, following burst: D correlation coefficient: 0.815 for the WT and 0.596 for the KO). All the correlations are summarized in Fig. \ref{corrFig}F. To conclude, the segmentation analysis can be used to study correlation between bursting events, suggesting here that possibly a control feedback mechanism could be lost in KO, resulting in refractory periods that are not long enough to allow the neurons to fully recover after a burst, therefore keeping them in a more depressed synaptic state.
\subsection{Deconvolution kernel between spikes recorded from different electrodes}
\begin{figure}[http!]
\centering
\includegraphics[scale=0.8]{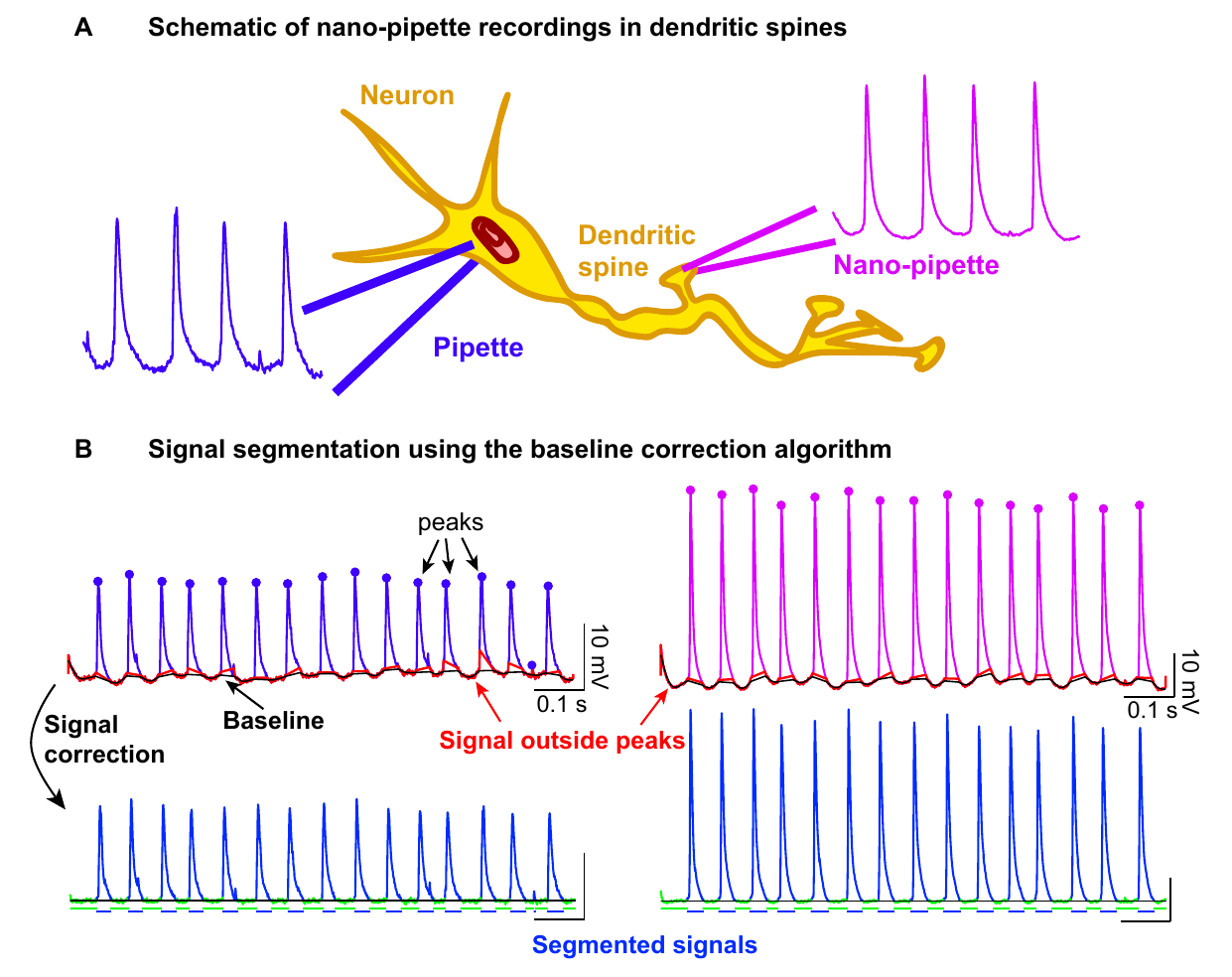}
\caption{{\bf Recording and segmenting spikes from normal and nano-pipettes.} {\bf A.} Schematic representation of dual recordings with pipettes of different sizes. {\bf B.} Pre-processing step to remove the low component followed by a segmentation to isolate the spikes.}\label{nanopip}
\end{figure}

\begin{figure}[http!]
\centering
\includegraphics[scale=0.8]{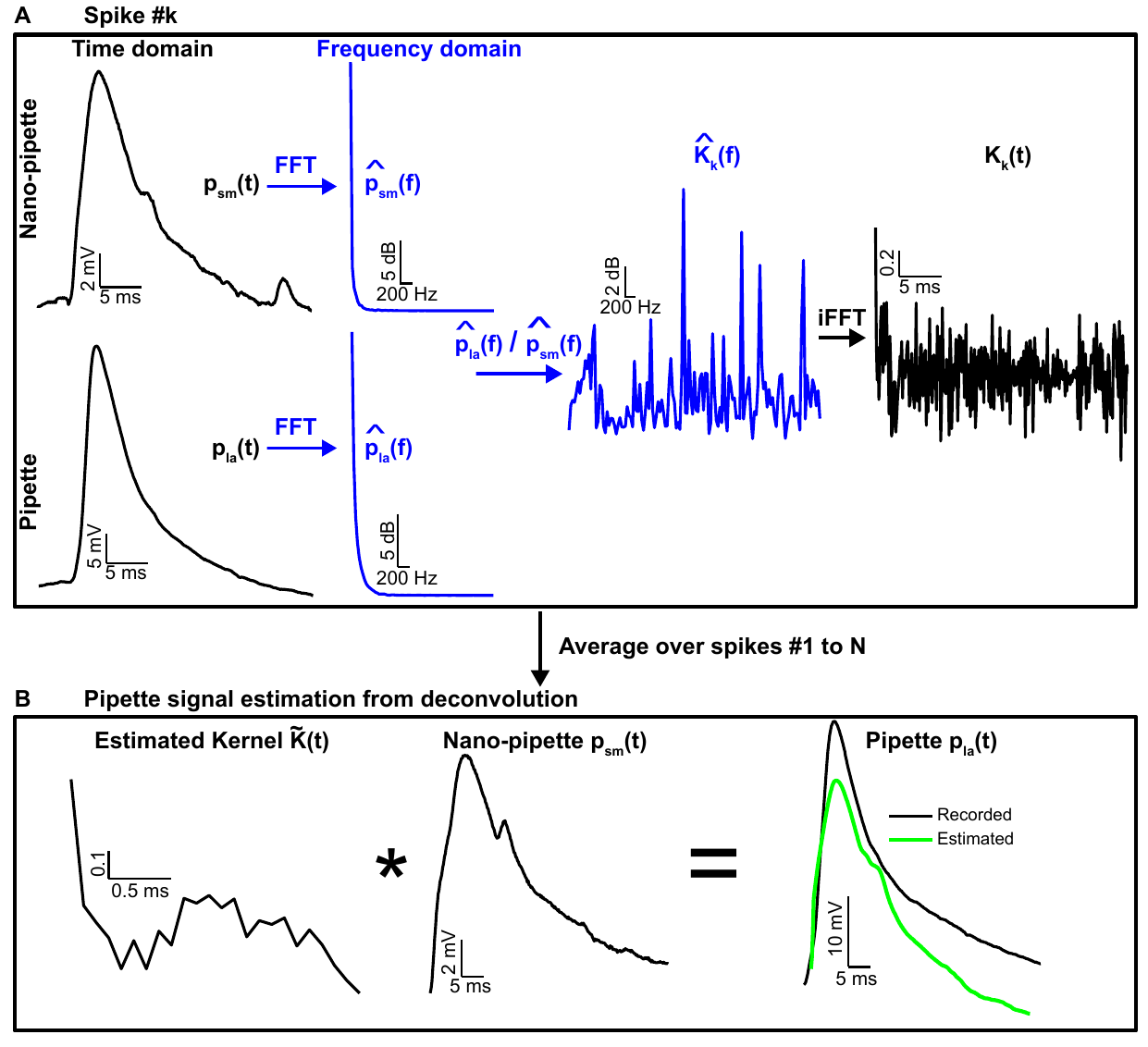}
\caption{{\bf Deconvolution of segmented spikes between normal and nano-pipettes.} {\bf A.} Estimation of the local convolution kernel $K_k$ from the pipette and nano-pipette signal during spike $k$. {\bf B.} Estimated kernel $\tilde{K}(t)$ obtained from averaging the spike 1 to $N$ (left). convolution with the a nano-pipette spike $p_{sm}(t)$ (center) allows to recover the pipette signal (right, green).}\label{nanopipConv}
\end{figure}

Another application of the continuous time segmentation and deconvolution methods presented above is to identify the relation between the amplitude and time responses of induced spikes between a normal pipette and a nano-pipette \cite{jayant2017targeted,jayant2019flexible} from recordings made on the same neuronal soma. Since a nano-pipette can access smaller regions, while the normal one cannot, a first step consists in calibration (Fig. \ref{nanopip}A) the small to the normal signal as the same electrophysiological signal should be reported by the two pipettes. We use a causal convolution transformation to map the response of the nano-pipette to the one of the normal pipette. This mapping requires first a pre-processing of the signal to remove the low component and thus align the baselines, which can be done using the method described in section \ref{s:calciumMethod} and applying the same segmentation procedure to isolate each spike (Fig. \ref{nanopip}B).\\
The deconvolution procedure consists in finding the kernel function $K(t)$ that sends the response $p_{sm}(t)$ of the nano-pipette to the response of the larger pipette $p_{la}(t)$, so that we have the causal relation
\beq \label{causal}
p_{la}(t)= \int_0^t K(t-s)p_{sm}(s)ds.
\eeq
The direct Fourier transform can be used here so that for each spike $k$, we get an approximated kernel $K_k$ obtained by inverting the Fourier transform of relation \ref{causal} (Fig. \ref{nanopipConv}A)
\beq
\hat{K}_k(f)=\frac{\hat{p}_{la}(f)}{\hat{p}_{sm}(f)}.
\eeq
We use the mean of all estimates as an approximation of a stationary kernel:
\beq
\tilde{K}(t)\sim \frac{1}{N} \sum_{1}^{N}{K}_k(t).
\eeq
In the case of a non-stationary signal, another possibility would be to average the last $N=10$ events before, so that the kernel can evolve with time. Finally, the inverse Fourier transform allows to recover the time-dependent kernel. \\
The method is validated on spikes that did not contribute to find the kernel (Fig. \ref{nanopipConv}B). Once the kernel is found, it can be used to deconvolve the signal recorded by the nano-pipette in nanoscale regions, where the large pipette cannot access. {The baseline correction and spike detection were performed using the calcium data analysis toolbox described in the previous section. The code for the deconvolution step and kernel reconstruction is provided here \href{www.bionewmetrics.org}{bionewmetrics.deconvolution}.}
\subsection{Conclusion about segmentation and toward the need of modeling for interpretation}
In the past two decades, automatized electrophysiological time series segmentation of long recordings has been used to reveal various type of voltage, timing and many more distributions. In addition, the focus of studying fluctuations of the membrane potential revealed how complex dynamics such as Up and Down states are associated with the first step of neuronal encoding of an external stimulation. When the neuronal tissue is studied in the limit where population burst emerges spontaneously, usually by switching off the inhibition, their dynamics can reveal some inherent properties of the underlying neuronal mechanism that could be prevalent in pathological conditions.\\
Collecting local transient duration distributions already revealed several physiological functions: 1) busting activity is more frequent but shorter in epileptic conditions after altering the astrocyte network \cite{Rouach_CxKO} compared to normal condition. 2) The similarity between Up states responses in vivo induced spontaneously or evoked suggest profound \cite{chen2013reactivation} mechanisms of brain representation and replay \cite{kenet2003spontaneously}.\\
Interpreting the statistical properties of bursting, duration of Up states, etc... is often difficult as they result from the combination of multiple mechanisms such as channel activation, neuronal connectivity and synaptic dynamics. The specific role of each of these components can be analyzed with a modeling approach, numerical simulations and also, when possible, by deriving the appropriate formulas that relate the multiple parameters together. \\
Furthermore, time responses or empirical distributions can provide serious constraints for the model ability to capture the appropriate physiological phenomenology. This is often revealed by the fitting step of analytical or simulated curves derived from the model. An optimal fit procedure should then be used to extract specific parameters of the model, as we shall explain in the next section.
\pagebreak
\section{Parsimonious neuronal network models based on short-term synaptic plasticity} \label{s:modeling}
A large class of models have been developed to account for neuronal networks activity, using channels dynamics \cite{hille1978} but also synaptic short-term properties \cite{tsodyks1998facilitation,tsodyks1998facilitation,Mongillo2008}.  These models can account for neuronal networks that exhibit a large variety of time-frequency behaviors during spontaneous or induced responses by external stimulations. The time-frequency transient responses are often complex as explored by the time-frequency spectrogram \cite{buzsaki2006rhythms} or correlation analysis. They can span a large frequency range, from 0 to hundreds of Hz, as well as time scales from a few milliseconds to several seconds.\\
Reduced models started with Hodgkin-Huxley \cite{HH} based on channel conductances, that summarize the neuronal activity as a single voltage dynamics representing an ensemble of connected neurons. Other models have followed with refined the dynamics according to the physiological properties they needed to account for. Morris-Lecar \cite{morrisLecar} and Li-Rinzel focus on calcium dynamics and many other variants are summarized in the book \cite{Izhikevitch2007}.\\
Interestingly, models that account for synaptic properties can already account for a large diversity of physiological behaviors such as synaptic reponses, Up-Down states \cite{Holcman_Tsodyks2006}, bursting dynamics \cite{daoduc2014frontiers,DaoDuc2015}, transient network response to an external stimulation \cite{Barak2007}, memory capacity retrieval \cite{mi2017synaptic} or theta-paced flickering between place-cell-maps in the hippocampus \cite{mark2017theta}. We present in this section coarse-grained models used to characterize the statistics of bursting, as discussed in the sections above. The models are based on mean-field approximation of neuronal network driven by synaptic short-term plasticity \cite{varela1997depressionModel}.
\subsection{Modeling short-term synaptic depression-facilitation between two neurons}
The initial neuronal synaptic model is based on the biophysical mechanism associated with short-term synaptic depression which results from a lack of releasable presynaptic vesicles. Lack of vesicles prevents the transmission between two connected neurons from occurring \cite{Tsodyks1997}. Specifically, each action potential (AP) of a presynaptic neuron triggers a vesicular release with a given probability, usually around $p\sim 0.2$. Thus, to describe the effect of an AP on the post-synaptic neuron, the change in the fraction of available resources (vesicles) is monitored by splitting the resources into three states: an effective state $E$ (vesicle are available), an inactive $I$ state (nothing is left) and a recovered $R$ state (which accounts for the time to reconstitute a pool of readily releasable vesicles) \cite{Tsodyks1997}. The equations associated to the synaptic response are
\beq
\arraycolsep=1.4pt\def\arraystretch{2.0}
\begin{array}{r c l}
\dfrac{dR}{dt} &=& \dfrac{I}{\tau_{rec}}-U_{SE}R\delta(t-t_{AP}) \\
\dfrac{dE}{dt} &=& \dfrac{E}{\tau_{inact}}+U_{SE}R\delta(t-t_{AP})\\
I&=&1-R-E,
\end{array}
\label{TMmodel}
\eeq
where $\tau_{rec}$ and $\tau_{inact}$ are the time scales for recovery and inactivation respectively and $U_{SE}$ reflects the fraction of used resources at each action potential occurring at time $t_{AP}$.
\subsection{Modeling short-term synaptic plasticity in a single neuron and at the network level}
The model equations \eqref{TMmodel} developed for a synaptic interaction can be extended to account for the dynamics of a homogeneously and sufficiently connected neuronal ensemble \cite{tsodyks1998facilitation}. In that case, the system becomes
\beq
\arraycolsep=1.4pt\def\arraystretch{2.0}
\begin{array}{r c l}
\tau \cfrac{dM}{dt} &=&-M + J x u M +I_{ext}(t)+ \sigma \dot{\xi}\\
\cfrac{dx}{dt} &=&\cfrac{1-x}{\tau_{D}} -u x M\\
\cfrac{du}{dt} &=&\cfrac{U-u}{\tau_{f}} +U(1-u)M,
\end{array}
\label{pavel}
\eeq
where the variable $M$ represents the mean firing rate of the neuronal population, the synaptic efficacy is modulated by the fraction of available synaptic vesicles $x$ and the release probability $u$ can be interpreted as presynaptic calcium accumulation \cite{DaoDuc2015}. The external input $I_{ext}(t)$ can vary in time and the additive noise $\sigma \dot{\xi}$ is approximated as a Gaussian.\\
The release probability $u$ is increased (facilitated) every time a spike is generated and increases the presynaptic firing rate, while synaptic resources decrease. In the absence of any presynaptic firing, the variables $x$, $u$ and $M$ recover to their resting values, 1, $U$ and 0 with time constants $\tau_{D}$, $\tau_{F}$ and $\tau$. The synaptic dynamics can be monitor by the product $Jux$. Instead of a linear neuronal input, the second term in the first equation of \eqref{pavel} can be replaced by $g(J x u M +I_{ext})$, where $g(z)=a \log (1+\exp z/b)$, where $a$ and $b$ are constants \cite{tsodyks2017}.
\subsection{Modeling short-term synaptic depression to account for Up and Down state dynamics}
\begin{figure}[http!]
\centering
\includegraphics[scale=0.8]{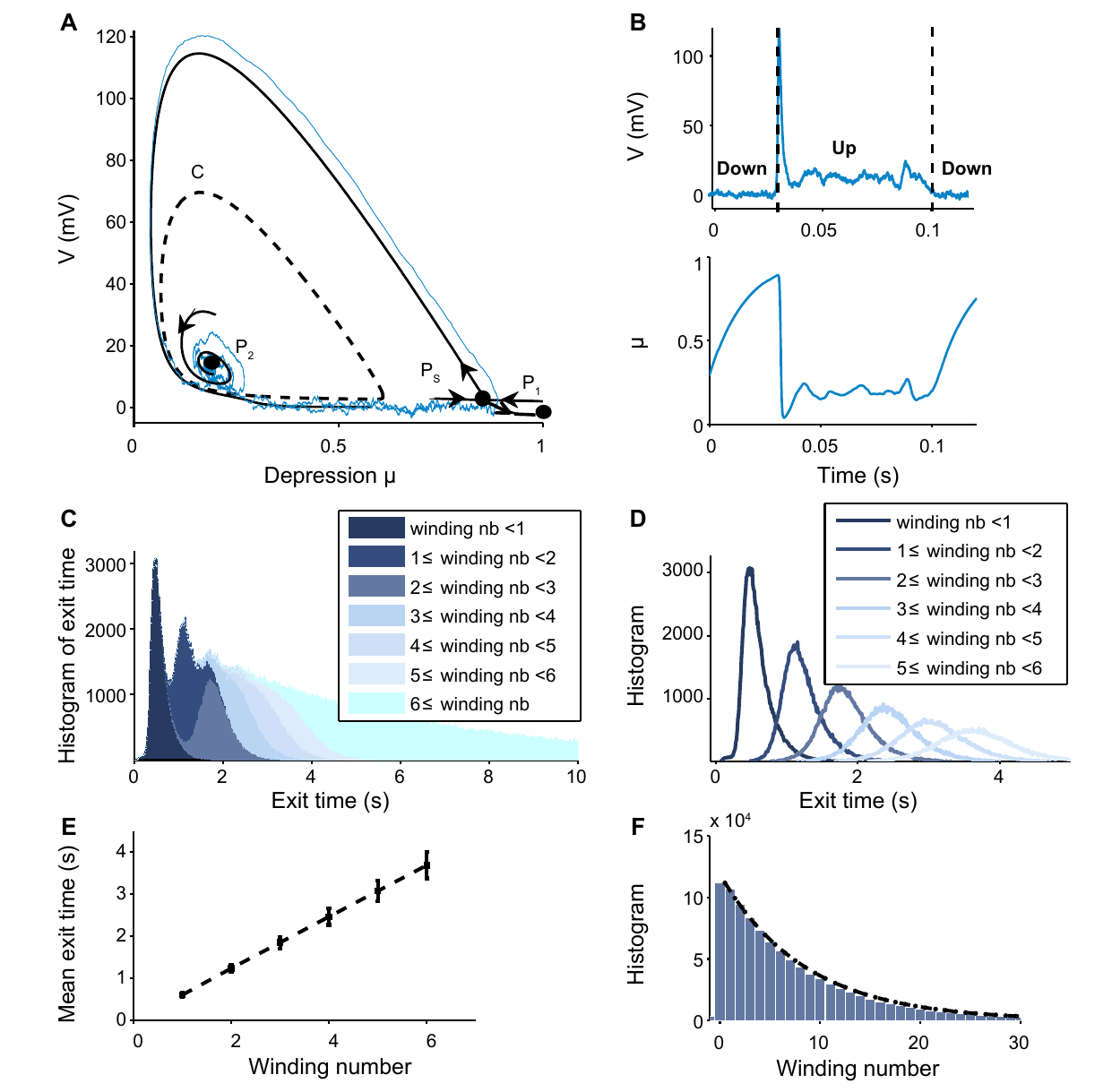}
\caption{Up-Down state model: {\bf A.} Two-dimensional phase-space of model \eqref{modelMMS} with two attractors ($P_1$ and $P_2$) and one saddle-point ($P_S$). Stochastic trajectories (blue) can escape from $P_1$ pushed by noise, they follow the unstable manifold of $P_S$ (black curve) and circle around $P_2$ until they cross the limit cycle (dotted line)and fall back to $P_1$. {\bf B.} Stochastic time-series of the two variables $V$ (upper) and $\mu$ (lower) exhibiting Up and Down states separated by vertical dotted lines. {\bf C.} Histogram of the exit times from the attractor point $P_2$ color coded by the number of revolutions around $P_2$ called winding number obtained from numerical simulations. {\bf D.}  Histogram of exit times from $P_2$ conditioned on the winding number. {\bf E.} Mean exit time as a function of the winding number from numerical simulations. {\bf F.} Distribution of the total winding number before exit.}\label{updownstate}
\end{figure}

Accounting for depression only, and thus keeping the facilitation constant in the mean-field firing rate eq. \ref{pavel},  with an additive Brownian noise leads to the simplified equations
\beq\label{modelMMS}
\arraycolsep=1.4pt\def\arraystretch{2.0}
\begin{array}{r c l}
\tau\dot{V}&=&-V+\mu U w_T \alpha (V-T)^+ + \sqrt{\tau}\sigma\dot{\omega}\\
\dot{\mu}&=&\dfrac{1-\mu}{t_r}-U\mu \alpha (V-T)^+,
\end{array}
\eeq
where $V$ is the average synaptic input (in mV), $(V-T)^+=max(V-T,0)$ the mean firing rate, $\alpha = 1Hz/mV$ a conversion factor, $w_T$ the average synaptic strength, $U$ the vesicle utilization parameter and $t_r$ the recovery time constant of the synaptic depression. Finally, $\dot{\omega}$ is the normalized white noise and $\sigma$ its amplitude. Under the condition that the synaptic weight is high enough $w_T>w_{c}$, where $w_{c}$ is a critical value above which the dynamics exhibits a bistability between two attractors (a focus and a center) separated by a separatrix and an unstable limit cycle (Fig. \ref{updownstate}). The stochastic dynamics allows to create random switches between these states. This dynamics recapitulates Up and Down states driven by short-term plasticity and synaptic noise in cortical neurons \cite{Holcman_Tsodyks2006}. The parameter values used in \cite{Holcman_Tsodyks2006} are summarized in Table \ref{table1}.
\begin{table}
\begin{center}
\begin{tabular}{|c|c|c|c|c|c|}
\hline
$\tau $ & $U$ & $J$ & $\sigma $ & $T$ & $t_{r}$ \\ \hline
$0.05$s &  $0.5$ & $12.6$mV/Hz & $2.2$mV & $2$mV & $0.8$s \\ \hline
\end{tabular}
\end{center}
\caption{Parameters of model \eqref{modelMMS}.}
\label{table1}
\end{table}

Analysis of model \eqref{modelMMS} reveals that the time distribution in Up states $f_{Up}$ is not Poissonian \cite{daoduc2016}, but consists in a sum of oscillating exponentials, but a truncation to the first two terms is acceptable:
\beq\label{SumOfExpFit}
f_{Up}(t) = \sum_{k\geq 0} e^{-\lambda_k t}\cos(\omega_k t + \phi_k) \approx Ae^{-\lambda_0 t} + Be^{-\lambda_1 t}\cos(\omega t + \phi)
\eeq
where $A, B$ are constants and the parameters $\lambda_0,\lambda_1, \omega, \phi$ can be related to the dynamics in the phase-space of the focus attractor in Fig. \ref{updownstate} \cite{daoducPRE}. A similar oscillation phenomenon was also reported in \cite{verechtchaguina2007,verechtchaguina2006_1,verechtchaguina2006_2}. For the fitting procedure and the range of used parameters, we refer to the procedure described below in section \ref{s:data} (see also Table \ref{table2}).
\begin{table}
\begin{center}
\begin{tabular}{|c|c|c|c|c|c|c|c|}
\hline
 Recordings & A     & B      & $\lambda_0$ & $\lambda_1$ & $\omega$ (Frequency) & $\phi$(Phase)\\ \hline
 Barrel cortex & 54.4438 & 13.100  & 1.92    & 1.99   & 31.41  & -3  \\ \hline
 Auditory cortex& 23.9135 & 10      & 1.48    & 1.87      &  21.22 & -4.3\\ \hline
\end{tabular}
\end{center}
\caption{Equation \eqref{SumOfExpFit} parameters.} \label{table2}
\end{table}
The analysis of this dynamics further reveals that during Up states, a dominant frequency emerges, which depends on the parameters. These oscillations can be located in the $\theta$ or $\alpha$ bands \cite{zonca2021alpha}, confirming that the Up-state activity generates oscillations, a process that thus could be simply explained by synaptic dynamics.
\subsection{Synaptic short-term depression with facilitation to model neuronal population bursts}
The previous system of equations \eqref{modelMMS} used to model depressed synapses associated in a well connected neuronal network is insufficient to account for bursting events, which usually require an amplification phase to propagate a single spiking event initiated by one cell into a population response. The facilitation property plays exactly such a role. System \eqref{pavel} can be extended by considering the three variables: the mean firing rate $h$, the facilitation $x$, and the depression $y$, reflecting the fraction of available vesicles. This new system satisfies
\beq
\arraycolsep=1.4pt\def\arraystretch{2.0}
\begin{array}{r c l}
\tau \dot{h} &=& - h + Jxy h^+ +I_{ext}(t) +\sqrt{\tau}\sigma \dot{\omega}\\
\dot{x} &=& \dfrac{X-x}{t_f} + K(1-x) h^+ \\
\dot{y} &=& \dfrac{1-y}{t_r} - L xy h^+ ,
\end{array}
\label{sysDaop}
\eeq
where $h^+ = max(h,0) $ is a linear threshold function of the synaptic current which represents the firing rate. The first equation describes the response of the network ensemble to an external stimulation $I_{ext}(t)$, the term $\dot \omega$ is an additive Gaussian noise and $\sigma$ its amplitude. This term represents fluctuations in the synaptic current (synaptic noise). The second equation describes the facilitation response, and the third one is the depression, similar to system \eqref{modelMMS}. The parameters $K$ and $L$ describe how the firing rate modulates the depression and facilitation variables. The time scales $t_f$ and $t_r$ define the recovery of a synapse from the network activity. Finally, the combined effect of synaptic short-term facilitation and depression on the mean voltage is described by the term $Jxy$ where the parameter $J$ represents the network connectivity (mean number of synapses per neuron) \cite{Bart}.\\
The dynamical system \eqref{sysDaop} has been used to study the bursting dynamics observed in an island of cultured neurons when inhibitory neurons are suppressed \cite{DaoDuc2015}. The model accounts for short-time memory effects: while a direct stimulation can generate a 2s burst, a second stimulation at 5s interval lead to a reduced 1s burst. However, when the second stimulation occurs 30s later, instead of a reduced second burst, the duration is back to 2s. What is the cause for such effect? can this model explains this property?\\
System \eqref{sysDaop} has been used to answer positively this question and to compute the bursting reverberation time $T_R$ \cite{DaoDuc2015}. The model could precisely account for this short-term adaptation and showed that the shortest second burst was the result of long recovery of the depression parameter: the second burst was initiated while the depression was still in the recovery phase after 5s, but not after 30s. Fitting the experimental data allowed to estimate the parameters reported in Table \ref{table1}.
\subsection{Estimation of the burst duration based on an approximated synaptic short-term depression-facilitation model}
Although model \eqref{sysDaop} is nonlinear, it can be used to estimate the bursting duration with respect to the model parameters, and in particular to the synaptic connectivity $J$. The first step consists in defining the bursting profile using the firing rate variable $h$: the duration of bursting is defined as the time during which the firing rare $h$ is above a threshold $h_{th}$. During this reverberation period, the firing rate remains approximately constant in the initial phase of the response, which allows a partial decoupling of the synaptic equations. The approximated system is
\beq
\tau \dot{h} &=& - h + Jxy h + H\delta(t-t_{stim})\notag\\
\dot{x} &=& \dfrac{X-x}{t_f} + K(1-x)H \label{sys}\\
\dot{y} &=& \dfrac{1-y}{t_r} - L xy H.\notag
\eeq
During the early bursting time, we approximate equations 2 and 3 of system \eqref{sys}, by considering the approximation $h(t) \approx H$ at its initial value, leading to this approximated system. This system can be used to estimate the burst duration $T_R$ with respect to the threshold $h_{th}$ by a direct linear integration.\\
Although this approximation could affect drastically the dynamics, it decouples the equations of system \eqref{sys} and by resolving them with respect to the variable $h$, lead to the approximation procedure (supplementary information of \cite{DaoDuc2015}). Using the notation $T_R = T_0 = \tau \ln \left(\cfrac{H}{h_{th}}\right)$ for $J=0$, the derived expression for the bursting time is
\beq
T_R (J) = \dfrac{1- JX\tau - \sqrt{(J\tau X -1)^2- 2J\tau X H(K-LX)\tau \ln \left( \cfrac{H}{h_{th}}\right)}}{J\tau X H (K-LX)}.
\label{TR1}
\eeq
Interestingly, comparing simulations with the plot of expression \eqref{TR1} shows that it is a good approximation for $J<1.9$. However, when $J> 1.9$, the reverberation time is larger than 1s. Thus, the approximation made in equation \eqref{sys} is not valid anymore. Other approximations are discussed in \cite{DaoDuc2015}. Interestingly, the formula reveals how each parameter contributes to the bursting time and the analysis further revealed a maximum for the bursting time: when the network is not enough connected, the bursting time increases with $J$ due to the facilitation that triggers a response in a large ensemble of neurons. However, in a too densely connected network, depression dominates, thus decreasing the burst duration.\\
Finally, it is possible to relate the abstract degree of synaptic connectivity $J$ used in equations \ref{sysDaop} to the actual number of synapses by calibrating the model with experimental data: this procedure has been applied in neuronal cultures, where the connectivity $J=40$ (for $\tau=0.05$) accounted for approximately 3000 synapses per neuron in hippocampal islands containing 5 to 30 neurons homogeneously connected \cite{DaoDuc2015}. Finally, the present analysis based on the depression-facilitation model differs significantly from fitting hidden-Markov model approach \cite{mcfarland2011explicit}, where several exponential distributions including the gamma, inverse Gaussian and geometric distributions are used to fit the experimental histograms of bursting time.\\
To conclude, the present coarse-grained neuronal model provides the basis to fit data and to extract parameters, as opposed to a fitting procedure based on machine learning, where the very large amount of parameters usually prohibit a biophysical interpretation.
\subsection{Modeling Burst and interbust dynamics using AfterHyperPolarization and short-term synaptic plasticity}
While the short-term synaptic plasticity model \eqref{sysDaop} could reproduce with a satisfactory precision the changes in the bursting patterns over tens of seconds, the model could not account for the long interburst durations. To account for long time scale and the transient return of the membrane potential to the resting potential, another mechanism had to be taken into account: indeed this long transient recovery of hundreds of milliseconds is known to result from the AfterHyperpolarization (AHP) mechanism. It is not mediated by synaptic properties, but rather by a combination of calcium and potassium channels with medium to slow timescales \cite{AHP_review}. While the bursting time depends on the reverberation events involving depletion of vesicular pools and the refilling time was determinant for the next burst generation, yet the membrane potential has its own independent dynamics. Thus, a new mechanism needed to be incorporated to account for bursting and AHP periods. With such mechanism, the model would then be able to predict not only bursting, interburst interval but also possible correlations that could occur between successive bursts, burst and inter-burst intervals and between successive inter-bursts.\\
Bursting events have been traditional modeled by several abstract variables in a phase space \cite{Izhikevitch2007}. Another possibility is to use Na$^+$ and delayed rectifier K$^+$ currents to produce basic spiking and add Ca$^{2+}$-dependent currents to generate bursting.  Here we will present an alternative approach that does not account for the direct channels biophysical properties but rather models the phenomenology of AHP, leading to hyperpolarization and a long transient return to equilibrium. Thus instead of adding the channel mechanisms, the effect of AHP directly on the short-term synaptic model \eqref{sysDaop} is taken into account \cite{AHPmodel}. This approach requires two parsimonious modifications to account for two features:
\begin{enumerate}
  \item Decreasing the mean voltage value after bursting. This effect is achieved by modifying the transient behavior: in the model \eqref{sysDaop}, a new equilibrium state is added, representing hyperpolarization (see below).
  \item Modifying the voltage timescales to account for the slower AHP dynamics recovery. Two new time constants have to be considered:  The slow $\tau_{sAHP}$ and the medium $\tau_{mAHP}$. During the period characterized by the medium scale $\tau_{mAHP}$, the voltage is forced to hyperpolarize. While the slow recovery to the resting membrane potential is dominated by the slow $\tau_{sAHP}$.
\end{enumerate}
The new coarse-grained AHP-short-term-plasticity model equations are summarized as follows:
\beq
\arraycolsep=1.4pt\def\arraystretch{2.0}
\begin{array}{r c l}
\tau_0 \dot{h}&=& -(h-T_0) + Jxy(h-T_0)^+ +\sqrt{\tau_0}\sigma \dot \omega\\
\dot{x} &=&\cfrac{X-x}{\tau_f} + K(1-x)(h-T_0)^+ \\
\dot{y} &=&\cfrac{1-y}{\tau_r} - Lxy(h-T_0)^+,\\
\end{array}
\label{AHP_model}
\eeq
where the variables $\tau_0$ and $T_0$ are now piecewise constant, depending on the burst decomposition into four phases, illustrated in Fig. \ref{model}A. We shall describe the four phases of a bursting-Interburst cycle within such a model.
\subsubsection{Step 1: Burst}
To emulate a population burst (Fig. \ref{model}A, blue) we use model \eqref{sysDaop}. A burst is associated with a decay of the depression variable $y$ and an increase of the facilitation variable $x$. These processes are governed by the fast time constant $\tau_0$ for the mean voltage variable $h$ which is $\tau_0=\tau$ while the resting value of $h$ is $T_0=T=0$ (see Table \ref{tableParam} for the parameters values). In practice, this behavior is defined for a certain configuration of the parameter space: $\{ y>Y_{AHP}$ and $h \geq H_{AHP}$ or $\dot{y} \leq 0 \}$ (Fig. \ref{model}B, above the purple and orange surfaces).
\subsubsection{Step 2: Hyperpolarization}
The hyperpolarization phase (Fig. \ref{model}A, orange) starts at the end of the burst as the depression variable $y$ reaches its minimum. In the model, this condition is represented in the equation by $y<\cfrac{1}{1+Lx(h-T_0)}$. To force the voltage to hyperpolarize, the resting value of $h$  has to change transiently to the value $T_0=T_{AHP}<T$. Finally, the time constant of $h$ changes to a medium value $\tau_0=\tau_{mAHP}$. This phase lasts as long as $y<Y_h$ (Fig. \ref{model}B, below the orange surface), which is a threshold value chosen empirically to provide enough time for the mean voltage variable $h$ to reach its maximum hyperpolarization value $T_{AHP}$.
\subsubsection{Step 3: Slow recovery period}
At the end of the hyperpolarization period, to account for the slow recovery phase of $h$ (Fig. \ref{model}A-B, purple), the time constant has to change to its slowest value $\tau_0=\tau_{sAHP}$. Finally, the mean voltage $h$ recovers from hyperpolarization by setting back the resting value to the initial value $T_0=0$. These two modifications allow the slow recovery of $h$ to its resting membrane potential. In practice, during this phase, the  variable $y$ is still increasing. To determine the end of this phase, each variable $h$ and $y$ must return close to their resting value, that is $\{Y_{AHP}<y$ or $h < H_{AHP}\}$.
\subsubsection{\bf Step 4: Quiescent phase (QP)}
\begin{figure}[h!]
\centering
\includegraphics[scale=0.8]{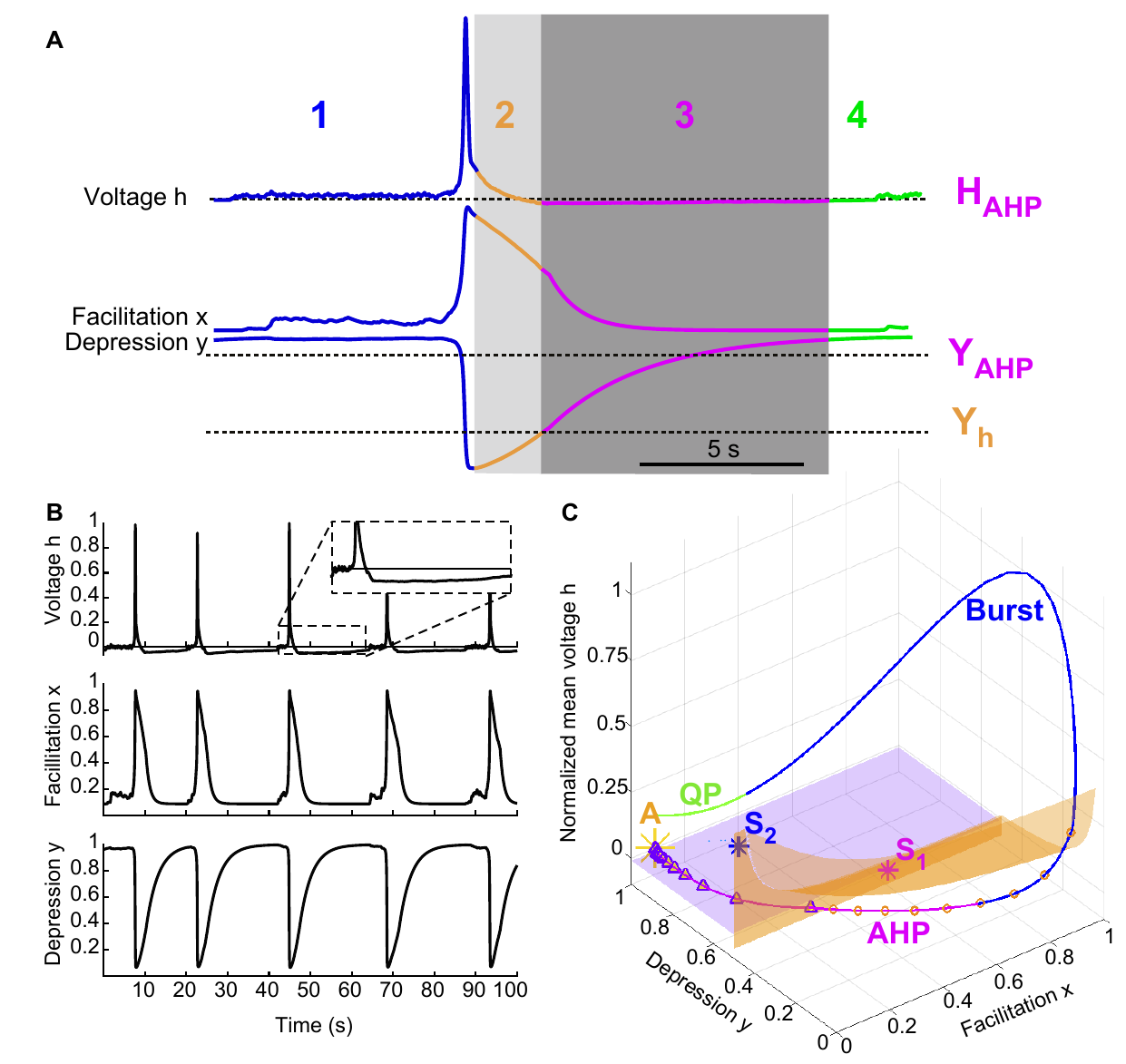}
\caption{\textbf{A.} $h$, $x$ and $y$ time-series decomposed over the four phases of a burst. \textbf{B.} Stochastic time-series of $h$, $x$ and $y$ exhibiting spontaneous bursts followed by AHP periods (inset). \textbf{C.} Phase-space of model \eqref{AHP_model} with a deterministic burst trajectory decomposed according to the four phases.}\label{model}
\end{figure}

Once system \ref{AHP_model} has recovered to a its steady state, the network is ready for a next bursting cycle. However, this occurs at a random time, triggered by fluctuations around equilibrium due to the random fluctuation (Fig. \ref{model}A-B, green). In this state, the parameter values are the ones of step 1: $\{y>Y_{AHP}$ and $h \geq H_{AHP}$ or $\dot{y} \leq 0\}$, $\tau_0=\tau$ and $T_0=T$.
\subsection{Accounting for neuron-glia interactions using AHP-synaptic short-term plasticity models}
Astrocytes form networks mediated by gap-junctions that interact and modulate the activity of neuronal ensembles \cite{giaume2010astroglial,dallerac2013astrocytes}. It remains quite difficult to model the interaction between neuronal and astrocyte networks because it involves ion exchanges with the extracellular space. For example, the dynamic of extracellular potassium levels has to be tightly controlled to avoid reaching concentrations that could generate epilepsy.\\
The facilitation-depression model with afterhyperpolarization model \ref{AHP_model}
can be used to model the consequences of the glial network alteration on the neuronal network by adjusting the key parameters underlying the bursting patterns regulated by the extracellular potassium. The parameters to be changed are the medium $\tau_{mAHP}$ and slow $\tau_{sAHP}$ time constants and the threshold $T_{AHP}$ that are defined by
\beq
\tau_0=
\begin{cases}
\tau_{mAHP} & \hbox{ if } y<Y_h \hbox{ and }\dot{y}>0 \\
\tau_{sAHP} & \hbox{ if } Y_0<y<Y_{AHP} \hbox{ and }h<H_{AHP}\\
\tau_0 & \hbox{otherwise}
\end{cases}
\eeq
and
\beq
T_0=
\begin{cases}
T_{AHP} & \hbox{ if } y<Y_h \hbox{ and }\dot{y}>0 \\
0 & \hbox{otherwise}
\end{cases}
\eeq
The adjustment procedure of these parameters is defined as follows: for each parameter set, the histograms of burst and interburst durations are generated by running numerical simulations and by segmenting the simulated time-series as described in subsection \ref{SegmentSim}. Then an optimization procedure is used to explore the parameter space and to determine the optimal set of parameters. This can be done on a wild type (WT) but also on a genetically perturbed mice.\\
Once the parameters are calibrated, this approach can be used to determine the relative contribution of the changes in either AHP, basal membrane depolarization, synaptic noise or depression to a connexin-deficient astroglial network (i.e. without gap junctions between the astrocytes). This approach allows to quantify the role of each of these parameters on the neuronal activity, by assessing its effect on the neuronal bursting pattern. \\
This paradigm is thus to run bursting simulations with parameters obtained in the fit of the WT distributions, except for some tested factor (AHP, noise amplitude or depolarization threshold), for which we can use the parameters obtained from the fit of a connexin deficient condition.
To conclude, simulations based on data reveals how a calibrated mean-field model can be used to identify the contribution of astrocytes on a neuronal network by controlling the AHP dynamics.
\subsection{Variant of the segmenting algorithm applied to simulated time-series}\label{SegmentSim}
A variant of the segmentation procedure of the burst and interburst interval adapted for the time-series generated by system \ref{AHP_model} is in contrast with the segmentation we presented for experimental time-series in subsection \ref{ss:segmentation}. Simulated time-series are generated from equations \ref{AHP_model} and the dynamics of the firing rate is driven by the small Brownian noise. Such a model can reproduce the spontaneous bursting dynamics, followed by long AHP refractory periods and a quiescent phase, characterized by local fluctuations of the firing rate around the resting state (Fig. \ref{model}C). The adjusted segmentation algorithm, leading to bursting, AHP and quiescent phase periods is as follows:
\begin{itemize}
  \item[-] {\bf Burst initiation:} A bursting event is first detected when the variable $h$ reaches a threshold $h(\tau_1)=T_1$, chosen sufficiently high to avoid detecting the small fluctuations due to the noise term. However, contrary electrophysiological time-series (section \ref{ss:segmentation}), where a burst initial rise is very fast, here we cannot neglect the time delay between the burst initiation and the time $\tau_1$ of burst detection. Thus, the time $\tau^{ini}$ of burst initiation is set equal to the last time previous to $\tau_1$, where the variable $h$ equals its resting value, that is $h(\tau^i)=T$ with $\tau^i<\tau_1$.
  \item[-] {\bf Burst termination:} the termination is detected by finding the time $\tau_2$ such that the variable $h$ passes a second threshold $T_2<T$ i.e. $h(\tau_2)=T_2$. We apply a similar correction to this time by taking the last time $\tau^e$ before $\tau_2$ where $h(\tau^e)=T$ with $\tau^e<\tau_2$.
  \item[-] {\bf AHP detection:} similar to the procedure used for patch-clamp recordings (section \ref{ss:segmentation}), the AHP period starts immediately at the end of the burst at time $\tau^e$, and lasts until time $\tau_e^a$ when the voltage $h$ reaches its equilibrium value, i.e. $h(\tau_e^a)=T$ and $\tau_e^a>\tau^e$.
  \item[-] {\bf QP detection:} The QP period starts at the AHP end time $\tau_e^a$ and lasts until the next burst is initiated.
\end{itemize}
We obtain the relevant statistics for burst, AHP and QP duration distribution by collecting the durations generated by the simulation based on segmentation procedure describe above. We shall use them in section \ref{paramEstimSection} to calibrate the model parameters, a necessary procedure to evaluate the role of each parameter separately. The code to simulated the model \eqref{AHP_model} and the segmentation  of the simulated time-series is available at  \href{www.bionewmetrics.org}{\url{bionewmetrics.ahp_model}}.
\begin{table}
\begin{center}
\begin{tabular}{|r|c|}
\hline
Burst detection threshold $T_{e1}$ & $100+T$ \\
End of burst \& AHP threshold $T_{e2}$ & $T-1$\\
\hline
\end{tabular}	
\end{center}	
\label{tableSegSim}
\caption{Simulated time-series segmentation parameters \eqref{modelMMS}.}
\end{table}
\begin{figure}[http!]
\centering
\includegraphics[scale=0.8]{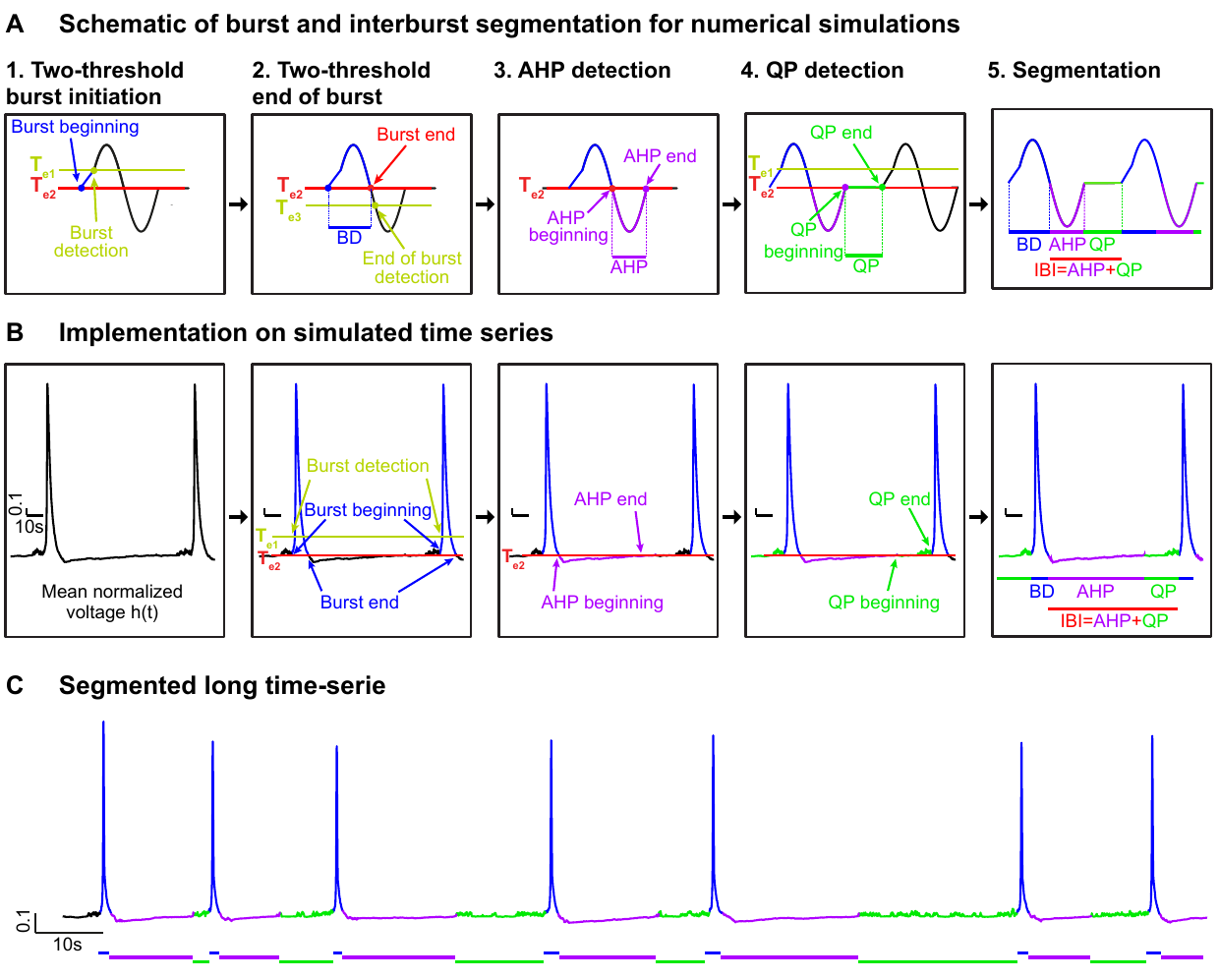}
\caption{\textbf{Segmentation pipeline for the simulated time-series} \textbf{A. Burst and interburst segmentation algorithm.} 1. Burst initiation is detected using the two thresholds $T_{e,1}$ (detection) and $T_{e,2}$ (real burst beginning) on the $h$ time-series. 2. End of burst is detected using the two thresholds $T_{e,3}$ (detection) and $T_{e,2}$ (real burst end) on $h$.  3. AHP begins at the end of the burst and lasts until $h$ is back above resting membrane potential ($=T_{e,2}$) 4. QP begins at the end of AHP until the next burst beginning. 5. Extraction of burst durations, (BD,blue), AHP duration (magenta) and quiescent phases (QP, green). IBI (red) is composed of AHP and QP. \textbf{B. Implementation on simulated time-series.} Step by step application of the algorithm described in A on simulated time-series. \textbf{C.} $200$s simulated time-serie of the mean normalized voltage $h$ segmented into burst (blue) AHP (magenta) and QP (green) periods using the algorithm described in A.}\label{simSeg}
\end{figure}
\subsection{Modeling slow rhythmic oscillations with two short-depression time scales for the preB\"otzinger complex}
\begin{figure}[http!]
\centering
\includegraphics[scale=0.65]{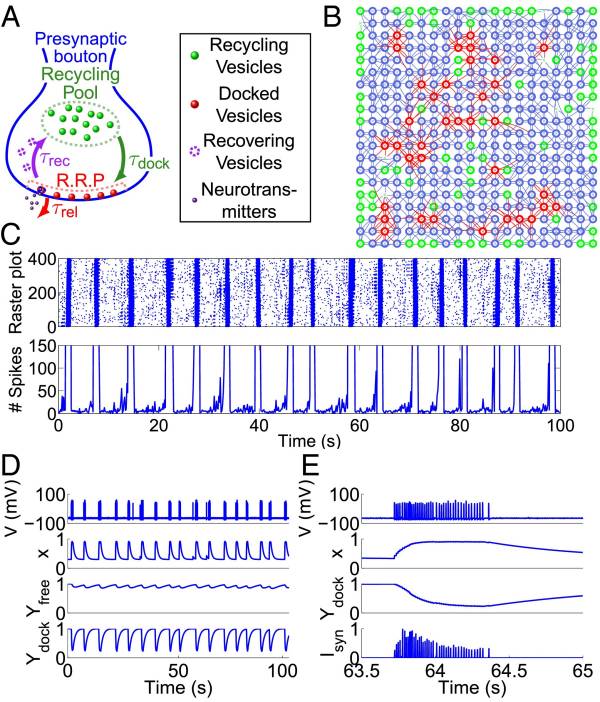}
\caption{{\bf PreBötzinger complex model and resulting network activity.} {\bf A.} Schematic representation of the presynaptic bouton: vesicles are divided into two pools: diffusing (green) and docked at the active zone (AZ, red) with recovering ones (purple). After fusion, vesicles recover and enter the recycling pool. {\bf B.} Example of a neuronal network (400 neurons) where neurons are located on a square lattice. The connections can be higher than 10 (red), between 5 and 10 (blue), or less than 5 (green). Neurons with no input or output are marked by green crosses, with no circles around. {\bf C.} (Upper) Raster plot for spiking neurons simulated during 100 s, the synchronous rhythmic patterns alternate between active and silent periods. (Lower) Time-dependent plot of the number of spikes in the network, computed in time windows of 100 ms. The $y$ axis is zoomed to observe the preburst increase in the number of spikes. {\bf D.} Time-dependent plot of the voltage $V$, the facilitation variable $x$, and the scaled variables $Y_{\text{free}}=y_{\text{free}}/y^{max}_{\text{free}}$, $Y_{\text{dock}}=y_{\text{dock}}/y^{max}_{\text{dock}}$ for a single neuron chosen randomly from the network (the mean bursting duration is $777 \pm 98$ ms, and the mean interburst interval is $5.2 \pm 1.0$ s). {\bf E.} Magnification of $V$, $x$, $Y_{\text{dock}}$, and $I_{syn}$ for the neuron in D, during 1.5 s (same simulation as C and D) copyright from \cite{Guerrier2015}.} \label{prebotz}
\end{figure}

Another variant of the depression-facilitation model has served to study the bursting activity generated by the preB\"otzinger complex (preBötC), which gives rise to the respiratory rhythm \cite{Feldman1991,Feldman2016,ashhad2022breathing}. The preB\"otC generates a spontaneous respiratory-like rhythm even in slices. The model developed in \cite{Guerrier2015} predicted that a neuronal network model based on two successive synaptic depression mechanisms (Fig. \ref{prebotz}A) allows a synchronized activity in the network (bursts) with a population refractory period (silence) that prevents an immediate bursting activity. In addition, combining electrophysiological recordings and the model suggested that the rhythm emerges through stochastic activation of intrinsic currents conveyed by recurrent synaptic connections (Fig. \ref{prebotz}B), without the need of rhythmic pacemaker neurons.\\
Bursting in this model (Fig. \ref{prebotz}C), is obtained by four principal state variables of a neuron (Fig. \ref{prebotz}D-E): the voltage $V$, the facilitation variable $x$, and the two normalized depression-related variables $Y_{free}$ and $Y_{dock}$. Synaptic fluctuations are the driving forces that can initiate population activities through recurrent excitation. The system of equation is given for each neuron $i$ by
\beq
\nonumber \displaystyle C \,\dot V_i&=&  -I_{Na}-I_{K}-I_{L}\displaystyle +\sum_{j \text{ connected to $i$}}I_{syn,j}+\sigma \dot W_i \\
 \displaystyle \dot x_i & = & \displaystyle \frac{X-x_i}{\tau_f} + k(1-x_i)H(V_i-T) \\
      \nonumber   \displaystyle \dot y_{\text{free},i}& = & \displaystyle \frac{1-y_{\text{free},i}-y_{\text{dock},i}}{\tau_{\text{rec}}}-\frac{1}
      {\tau_{\text{dock}}}\left(y_{\text{dock}}^{\max}-y_{\text{dock}}\right) y_{\text{free},i} \left[1+\frac{x_i-X}{X}H(V-T) \right] \\
   \nonumber     \displaystyle \dot y_{\text{dock}} & = & \displaystyle \frac{1}{\tau_{\text{dock}}}\left(y_{\text{dock}}^{\max}-y_{\text{dock},i}\right) y_{\text{free},i} \left[1+\frac{x_i-X}{X}H(V_i-T) \right]\\
    \nonumber  & & -\frac{1}{\tau_{\text{rel}}} y_{\text{dock}} \frac{x_i-X}{X}H(V_i-T),
\eeq
where the variables and parameters are described below. When a presynaptic AP arrives at a time $t_0$, which does not fall into the refractory period window, the synaptic current is
\beq
i_{t_0}(t) = K_I (y_{\text{dock}}(t)-y_{\text{dock}}(t_0))H(V-T)H\left(y_{\text{dock}}-y_{\text{dock}}^{\min}\right),
\eeq
where $K_I$ is a constant, which converts the variable accounting for the proportion of fused vesicles during an AP, into a post-synaptic current. For spikes arriving at times $t_k$, the synaptic current is for $t \in [t_k, t_{k+1}]$,
\begin{align}
   I_{syn}(t) = \begin{cases}
                    0 \;\;\text{during a refractory period}\\
                    i_{t_k}(t-\tau_{del}),\;\; \text{else}
                \end{cases}
\end{align}
where the rest of the currents are given by the classical Hodgkin-Huxley model \cite{hille1978}
\beqq
 I_{Na} &=& g_{Na}m_\infty^3h(V-E_{Na})\\
 I_K &=& g_{k}n^4(V-E_K)\\
 I_L &=& g_L(V-E_{L})\\
 m_\infty &=& \frac{\alpha_m}{\alpha_m+\beta_m} \\
\alpha_k &=& \frac{1}{\tau_k}\frac{\theta_k-V}{e^{\frac{\theta_k-V}{\tau_k}}-1} \hbox{ for } k=n,m \\
 \beta_k &=& \eta_ke^{-\frac{V+65}{\sigma_k}}.
\eeqq
The variables $m$ and $n$ represent the opening of the $Na^+$ and $K^+$ channels respectively. For the closing of the $Na^+$ channel, $h =(0.89-1.1n)$.
The synaptic current $\ds{ \sum_{j } I_{syn,j}}$ integrates the sum over all connecting neurons \cite{Guerrier2015}.\\
The synaptic dynamics depends on a two pool model that accounted for two different time scales of depression. Indeed, synaptic depression results from the depletion of the readily releasable pool (RRP) (Fig. \ref{prebotz}A), where synaptic vesicles are gathered at the membrane before fusion. The other pool of recycling vesicles (recycling pool, RP) are diffusing. Finally after fusion, vesicles are not participating in any of the two previous pools. This state is described as recovering. In the present model, the total amount of vesicles in a synapse is constant. Thus the fraction $y_{\text{free}}$ (resp. $y_{\text{dock}}$) of vesicles in the RP (resp. RRP), with the fraction of recovering vesicles $y_{\text{rec}}$ satisfies the conservation equation
\beq
\ds y_{\text{free}}+y_{\text{dock}}+y_{\text{rec}}=1.
 \label{eqvesicle}
\eeq
Finally, the synaptic facilitation variable $x$ reflects all possible mechanisms that enhance vesicular release, and thus is associated with an increase in the release probability \cite{Tsodyks1997}. It was already described in equation \eqref{sysDaop} where $\tau_f$ is the facilitation rate, $X$ its value at equilibrium, and $H$ is the Heaviside function. The facilitation $x$ increases due to the term $k(1-x)$ during an AP, when the membrane potential $V$ is above a threshold $T$ ($H(V-T) =1$), and relaxes back to equilibrium when $V$ is below $T$ ($H(V-T) =0$).\\
Simulations of the model revealed that for the chosen set of parameters, the fraction of vesicles in the RP ($Y_{\text{free}}$) is quite stable, oscillating between 80\% and 98\% of its maximum value, whereas the fraction of docked vesicles in the RRP ($Y_{\text{dock}}$) fluctuates between 20\% and 100\% of its filled state. The simulations suggested that the RRP alternates between an empty and full state, where the mean maximum number of vesicles is 4.5. When the percentage of vesicles in the RRP $Y_{\text{dock}}$ reaches its minimal value, the bursting period ends, leading to a decrease in the facilitation variable $x$ that relaxes back to equilibrium (Fig. \ref{prebotz}E). On the contrary, $x$ increases exponentially when each population burst begins, which, associated with the decrease in $Y_{\text{dock}}$, is reflected by a transient increase and subsequent decrease in the amplitude of synaptic currents during the bursts ($I_{syn}$, Fig. \ref{prebotz}E). \\
Each neuron within the network that reaches its minimum $y^{min}_{\text{dock}}$ enters a refractory state that shuts down its synaptic transmission. This minimum value leads to burst termination across the population because recurrent excitation ceases. As synapses recover, a subsequent burst can begin when several connected neurons spike in a short time interval to facilitate postsynaptic neuronal activation. This nonlinearly promotes spiking activity to spread through neighboring neurons and invade the entire network. These features were stable with respect to the network topology with a cycle periods (CPs), burst durations (BDs), and interburst intervals (IBIs) of $5.5 \pm 1$ s, $694 \pm 138$ ms, and $4.8 \pm 1$ s, respectively ($n = 10$ networks), which is in the range observed in vitro. \\
To conclude, this model could account for a large range of in vitro and in vivo data as well as small events called small burstlets, that could be observed before the emergence of a large burst \cite{cui2016defining}. The codes used in this project are available for download at \href{https://bionewmetrics.org/robust-network-oscillations-during-mammalian-respiratory-rhythm-generation-driven-by-synaptic-dynamics/}{\url{bionewmetrics.preBotz}}.
\subsection{Short-term plasticity model predicts the capacity of working memory}
Another application of synaptic short-term plasticity neuronal network concerns the capacity of working memory. How many random tasks or words can be transiently stored in the human memory? This number is related to the capacity of working memory (WM) \cite{mi2017synaptic}, limited for most human to four.\\
To study the neuronal computation underlying such a number, a network model was proposed in \cite{mi2017synaptic}, consisting in combining excitatory networks with synaptic currents. The mean-field short-term plasticity model consists in combining $P$ excitatory networks, as presented in equations \eqref{sys} with a common inhibitory one, that could inhibit all of them simultaneously. Using equations \eqref{sysDaop}, for $\mu=1..P$, the model can be written as
\beq\label{mishaeqs}
\arraycolsep=1.4pt\def\arraystretch{2.0}
\begin{array}{r c l}
\tau \cfrac{dh_{\mu}}{dt} &=& -h_{\mu}+J_{EE}u_\mu x_{\mu} R_{\mu}-J_{EI}RI+I_b+I_e(t),\\
\cfrac{du_{\mu}}{dt} &=& \cfrac{U-u_\mu}{ \tau_f}+U(1-u_\mu)R_\mu,\\
\cfrac{dx_{\mu}}{dt} &=& \cfrac{1-x_\mu}{\tau_d}-u\mu x_{\mu}R_{\mu},\\
\tau \cfrac{dh_I}{dt} &=& -h_I+J_{IE} \sum_{\nu} R_{\nu},
\end{array}
\eeq
where the variables are the firing rate $h_I$ of the inhibitory network, the time constant $\tau$ of excitatory and inhibitory neurons. $h$ and $R$ are the synaptic current and firing rate of excitatory and inhibitory neurons, respectively. The conversion to firing rate uses the function $R(h) = \alpha \ln(1+\exp(h\alpha))$, which is a neuronal gain chosen in the form of a smoothed threshold-linear function. \\
The current $I_b$ represents the background excitation input and $I_e$ is the external input. As described in the previous subsections, $u$ and $x$ refer to the
short-term facilitation and depression effects, respectively. The neuronal connectivities are encoded in the parameters $J_{EE}, J_{EI} $ and $J_{IE}$ that should be tuned.\\
When several items are generated by applying transient external excitation inputs to the several networks, a sub-number can be maintained successfully in the form of brief reactivation called population spikes. This number is around 4 and thus the interpretation is that the capacity of WM of four (see figure 1 of \cite{mi2017synaptic}).\\
The model allowed to estimate the capacity number $N_c$ as the ratio of the duration $T_{max}$ of the cycle between two consecutive recalls for a given network to the total duration $t_s$ of separation between population spike generated by all neuronal networks. \\
The cycle duration can be approximated as the time it takes for the synaptic efficacy curve to reach a peak while neglecting the firing rate, thus from equations 2 and 3 of the system \ref{mishaeqs}, we have  $T_{max}=\tau_d\frac{\tau_f/\tau_d}{1-U}$. The duration $t_s$ depends on 1) the width of the population spike of the previous item, the delay and the width of the inhibitory pulse triggered by this population spike, and the time for a next cluster to recover from inhibition and initiate a new population spike. A refined analysis leads to the expression $t_s= \tau \ln (\frac{h_0}{I_b-I_c})+C$, where $C$ is a constant, the negative initial value $h_0$ is determined by the strength of inhibition triggered by a population spike of a previously activated network and $I_c$ is a critical value parameter. This leads to the number of memory items
\beq
N_c=\frac{\tau_d\cfrac{\tau_f/\tau_d}{1-U}}{\tau \ln \left(\cfrac{h_0}{I_b-I_c}\right)+C}.
\eeq
To conclude, an elementary network configuration exhibiting short-term depression and facilitation where all excitatory networks are connected to the same inhibitory network can generate a memory replay where the mean capacity is $N_c$. This number depends on the intrinsic time scale of the synaptic dynamics as well as the amplitude of some input currents.
\subsection{Coupling synaptic short-term plasticity neuronal networks to generate brain oscillations}
\begin{figure}[http!]
\centering
\includegraphics[scale=0.7]{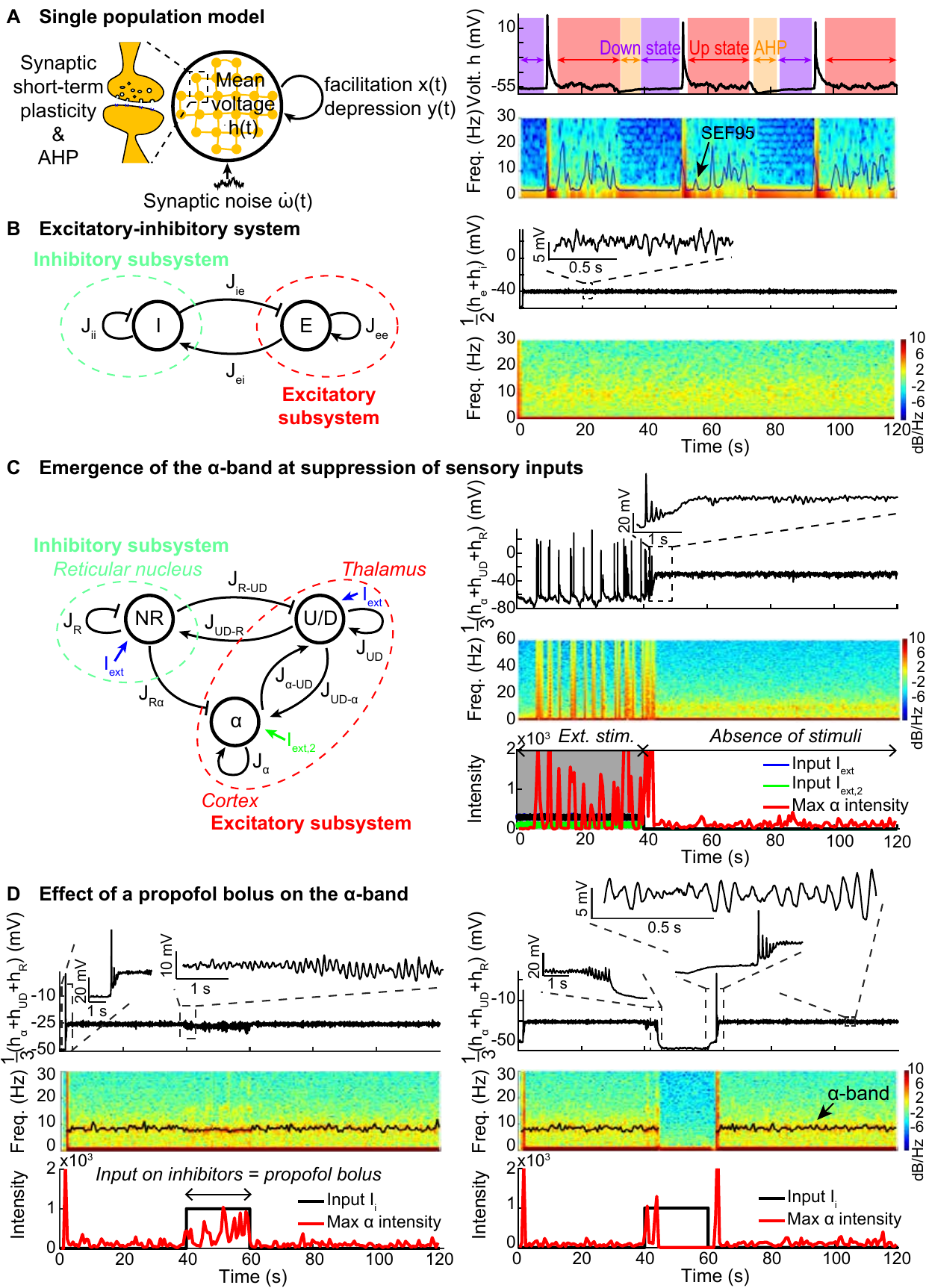}
\caption{\textbf{Modeling oscillatory rhythms using 3 network models.} {\bf A.} Single population with short-term facilitation-depression and AHP (left) producing slow oscillations between Up (red) and Dow (purple) states (right). {\bf B.} Excitatory population connected to an inhibitory one (left) generating a fragmented $\alpha$-band (right) {\bf C.} 3 networks: Up-down state oscillator ($U/D$) and Up-state locked excitatory populations ($\alpha$) both connected to an inhibitory network ($NR$, left), generating a complex pattern: bursting with external stimulation and stable $\alpha$-oscillation in the absence of stimulation, modeling the maintenance phase of general anesthesia. {\bf D.} Increase of inhibition generating a small decay of the main $\alpha$-frequency (left) or creating a complete suppression (right) adapted from \cite{zonca2021alpha}.} \label{AnesthesiaFig}
\end{figure}

Brain rhythms observed in EEG recordings often originate from multiple neuronal projections among different regions. For example the $\alpha$-band [8-12]Hz, observed in the prefrontal cortex during general anesthesia \cite{bacsar2012short,bacsar2012review}, involves thalamo-cortical projections \cite{dossi1992electrophysiology,steriade1994cortical, contreras1995cellular, steriade1996synchronization}.  We discuss briefly in this subsection how the $\alpha$ and the $\delta$ bands can be generated by connecting different neuronal networks characterized by depression-facilitation when all external sensory stimuli are shut down (Fig. \ref{AnesthesiaFig}A-B) \cite{zonca2021alpha}.\\
General anesthesia on the propofol anesthetic is characterized by a dominant $\alpha$-band but also low frequency delta-oscillations resulting from the switching between Up and Down states ($\delta$-band). In the Up-state, the model has to reproduce the $\alpha$-oscillation, as possible using equation \eqref{modelMMS}, the dynamics of which produce a focus attractor \cite{DaoDuc2015,daoduc2016}.  Indeed, the dynamics is driven by random fluctuations of the membrane potential. Switching between Up and Down state is associated with the slow $\delta$-band oscillation. These oscillations can be reproduced by adding AHP to short-term synaptic plasticity with the use of equations \eqref{AHP_model} (Fig. \ref{AnesthesiaFig}A). In this case, the overall network connectivity modulates the time spent in the Up-state.\\
When an inhibitory neuronal network is connected to an excitatory one with AHP, with a constant input current on the inhibitory population, then the network exhibits switching between Up and Down states. Furthermore, the injected current modulates the fraction of time spent in Up vs Down states. When this input current reaches a threshold, the excitatory population can be locked in a Down state, silencing the network. However, without AHP the inhibitory network cannot induce switching between Up and Down states, and the dominant $\alpha$-band generated by the excitatory population remains locked in the Up state (Fig. \ref{AnesthesiaFig}B, right).\\
By further coupling two excitatory networks with an inhibitory one (Fig. \ref{AnesthesiaFig}C, left), $\alpha$-oscillations are generated associated with a slow switching between Up and Down states ($\delta$-band) thus reproducing the variety of behaviors that can be generated by the thalamo-cortical loop: the excitatory subsystem contains two populations: 1) an $\alpha$-oscillation generator and 2) AHP-short-term plasticity network that generates Up-Down state switching. These two populations are connected by reciprocal connections and receive an inhibitory input from the inhibitory subsystem (Fig. \ref{AnesthesiaFig}C, left). This inhibitory network can also be activated by an external stimulation $I_i$ that models the injection of propofol.\\
To conclude, by connecting three neuronal populations, Up and Down states can coexist with an $\alpha$-band \cite{zonca2021alpha}: a stable $\alpha$-band emerges once external stimuli are suppressed by the anesthetic agent (Fig. \ref{AnesthesiaFig}B, right). This behavior results from the interactions between the three networks. The amplitude of inhibitory inputs as well as the connectivity level are fundamental parameters that define the proportion of Up states and thus the persistence of the $\alpha$-oscillations (Fig. \ref{AnesthesiaFig}D). The associated equations are presented and discussed in \cite{zonca2021alpha}. The codes are available  \href{https://zenodo.org/record/5708312#.YqxcUOzP2F4}{\url{zenodo.zonca.codes}}. These codes allow to simulate time series for the one, two and three population models where the connectivity matrices and other parameters can be adjusted as well as to detect the Up and Down states and track the $\alpha$-band in the simulated data. This modeling approach shows that fundamental brain rhythms can result from short-term synaptic plasticity interacting neuronal networks.
\section{Data modeling, parameters estimation, validation and predictions}\label{s:data}
In the previous sections, we presented various modeling approaches that can reproduce neuronal network activity based on synaptic plasticity. However, to interpret specific patterns, such models need to be calibrated to data and this procedure is not universal. We discuss here some possible calibrations based on comparing statistics obtained from the experimental data with the ones generated by numerical simulations. A well calibrated model allows to explore the physiology in various conditions and to comprehend or to dissect the mechanism underlying a given behavior. Models can also selectively explore the role of any relevant parameter.
\subsection{Model parameters estimation}\label{paramEstimSection}
Parameters estimation is a fundamental step to connect a computational model to empirical data. However there are no general ground truth principle for such a task and thus we are still lacking a common routine methodology. In the past decades, a model that was able to reproduce qualitatively the trends was considered to be satisfactory enough. Today, with refined modeling and data processing approaches, not only do we expect the models to be derived from fundamental biophysical of physiological principles, but also, parameters have to be estimated accurately. We expect that a given set of parameters can account for a diversity of observed physiological responses. Thus a robust model with a given parameter set should reproduce normal behavior and even behavior in modified conditions such as the ones induced when a specific physiological pathway is disrupted for example during a gene deletion experiment.
\subsection{Lack of a principle in parameter estimation procedures}
\subsubsection{Small dimensional space of parameters}
For a model with few parameters, known as low dimensional space, it is satisfactory to define a set of admissible parameters based on comparing mean and variance between simulations and experimental data. For modeling the bursting time, fixing the mean and variance of the response at 5 and 35 s was sufficient to determine the entire parameter space (Fig. \ref{reverberation}) \cite{DaoDuc2015}. \\
We shall now describe this case in more detail: there is a temporal correlation between bursts generated by small island of neurons \cite{Cohen_Segal2009}: a burst can be induced by a single neuronal stimulation that reverberate during few seconds into the ensemble of interconnected neurons (Fig. \ref{reverberation}). Interestingly, the same stimulation generated few seconds after the first one lead to a decrease of almost half the duration of the first burst, while another stimulation generated 30 s later triggers a burst of duration comparable to the first one. The modeling approach revealed that such bursting temporal correlation is due to the short-term synaptic depression property, as confirmed experimentally by blocking the synaptic transmission \cite{DaoDuc2015}.\\
To do so, the model has to be calibrated to reproduce the mean and the variance of the of burst dynamics as well as the decrease in the duration of the second burst. Specifically, the model contains 6 parameters, 4 were already known, leaving 2 to be calibrated. These constraints are quite strong and such success in reproducing the observed bursting dynamics suggested that the model captures the main mechanism underlying these bursting events. In particular, the long recovery timescale of the depression $\tau_r \approx 3s$, could explain that it takes tens of seconds for the synapse to be reactivated after a burst. Interestingly, the same model was used to reproduce the same phenomenology in hippocampal slices instead of islands by increasing further the depression timescale to $\tau_r \approx 20s$. Finally, having a calibrated model allows to explore the conditions of neuronal network physiology. In particular, the model revealed that the physiology conditions are exactly those that sets the longest burst duration. Further, as the mean number of synapses increases, which could reproduce development condition, the burst duration passes by a maximum before decreasing.
\subsubsection{Large dimensional space of parameters}
When a model contains a larger number of parameters (more than ~10), stringent criteria are needed to determine the parameter values. This can be achieved by comparing distributions instead of restricting to first and second statistical moments. In physiological models, parameters have a specific interpretation, which is not necessarily the case for machine-learning approaches, where the number of parameters have an other order of magnitude, up to millions \cite{gomez2021deepimagej}. The goal of a large parameter set is often to perform pattern recognition, the goal of which is different from studying the mechanism underlying a physiological function.\\
We discuss now a possible unified method to calibrate model parameters. In principle, this task is model dependent because each model has a specific purpose depending on the spatial and temporal scales it has been elaborated for. For mean-field network models, the goal is often to study neuronal properties from elementary processes such as synaptic-short term plasticity or channels behavior. For these models, the behavior or phase-space often depends on the range of the parameters. Thus finding the appropriate parameters is critical to obtain the relevant behavior and thus to clarify the complex dynamics.\\
For example, to obtain the emergence of Up and Down states with the synaptic depression equations \eqref{modelMMS}, the synaptic strength parameter $J$, associated with the average number of synaptic connections has to be large enough \cite{Holcman_Tsodyks2006}. This qualitative description is certainly not sufficient nowadays and a comparison to data is requested. This demands can be answered by introducing a comparison criteria: for example, can the model or generated distribution predict the distribution of times in the Up states compared to other models \cite{DaoDuc2015}? The model equation \ref{sys} \cite{daoduc2016} predicts that the distribution does not follow a Poissonian or Gamma distribution, but rather could be explained by a sum of oscillating exponentials. This oscillating behavior is obtained by the probability distribution function derived from the associated Fokker-Planck equations \cite{daoduc2016,DaoDuc2015,daoducPRE}. Indeed, the distribution of times in the Up state is approximated as
\beq
f_{Up}(t) =A \exp (-\lambda_0 t) +B  \exp (-\lambda_1 t) \cos(\omega_1 t +\phi_1).
\eeq
This distribution contains 6 parameters: $A, B, \lambda_0, \lambda_1, \omega_1, \phi_1$, with $\lambda_1>\lambda_0$. The constants $A$ and $B$ represent the relative contribution of the two main exponentials. The main decay is obtained by fitting the distribution of times in the Up state by the first exponential, leading to a value for the parameter $\lambda_0$. The distance between the two consecutive peaks is equal to $\cfrac{2\pi}{\omega_1}$, while the phase $\phi_1$ accounts for a possible shift in the position of the maximum \cite{daoduc2016}. Finally, the second exponential decay term $\lambda_1$ can be found by multiplying the empirical distribution by $\exp(\lambda_0 t)$ and then by fitting the decay tail by an exponential. To conclude short-term synaptic models are used to derive curves to fit time distributions (Fig. \ref{Updownsegmentation}).
\subsection{Identifying parameters using distributions} \label{pipelinemeth}
\begin{figure}[http!]
\centering
\includegraphics[scale=0.8]{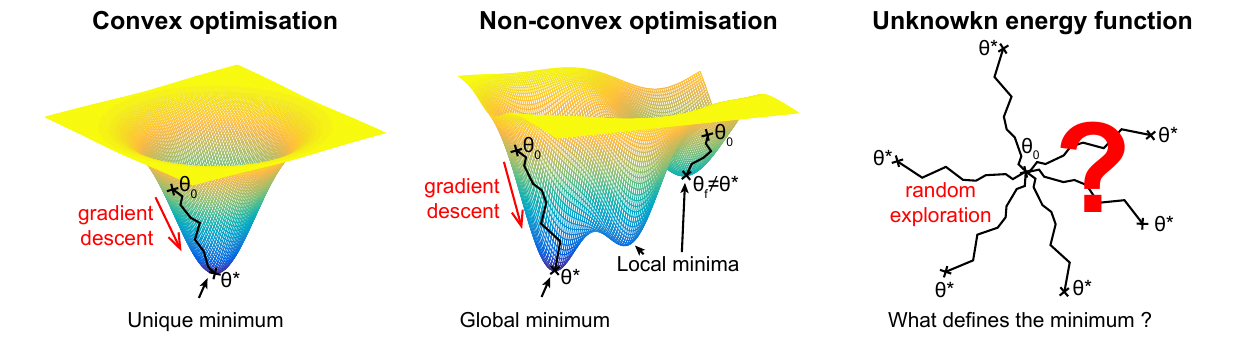}
\caption{\textbf{Optimization Procedures.} Illustration of an heuristic to select the optimal parameters: {\bf Left.} Convex optimization allows to find iteratively the optimal parameter values by converging to the global minimum $\theta^*$ of an energy function (yellow surface) starting from any point $\theta_0$ of the parameter space. The fastest path follows the steepest slope using a gradient descent algorithm (red arrow). {\bf Middle.} Non-convex optimization can lead to multiple local minima ($\theta_f \neq \theta^*$) and thus the selection of the optimal parameters depends on the initial point. {\bf Right.} When no energy functional is known, there are  standard procedure to  known to find the best parameter values. A possible approach is to use a random exploration with multiple scales.} \label{optimFig}
\end{figure}

\begin{figure}[http!]
\centering
 \includegraphics[scale=0.8]{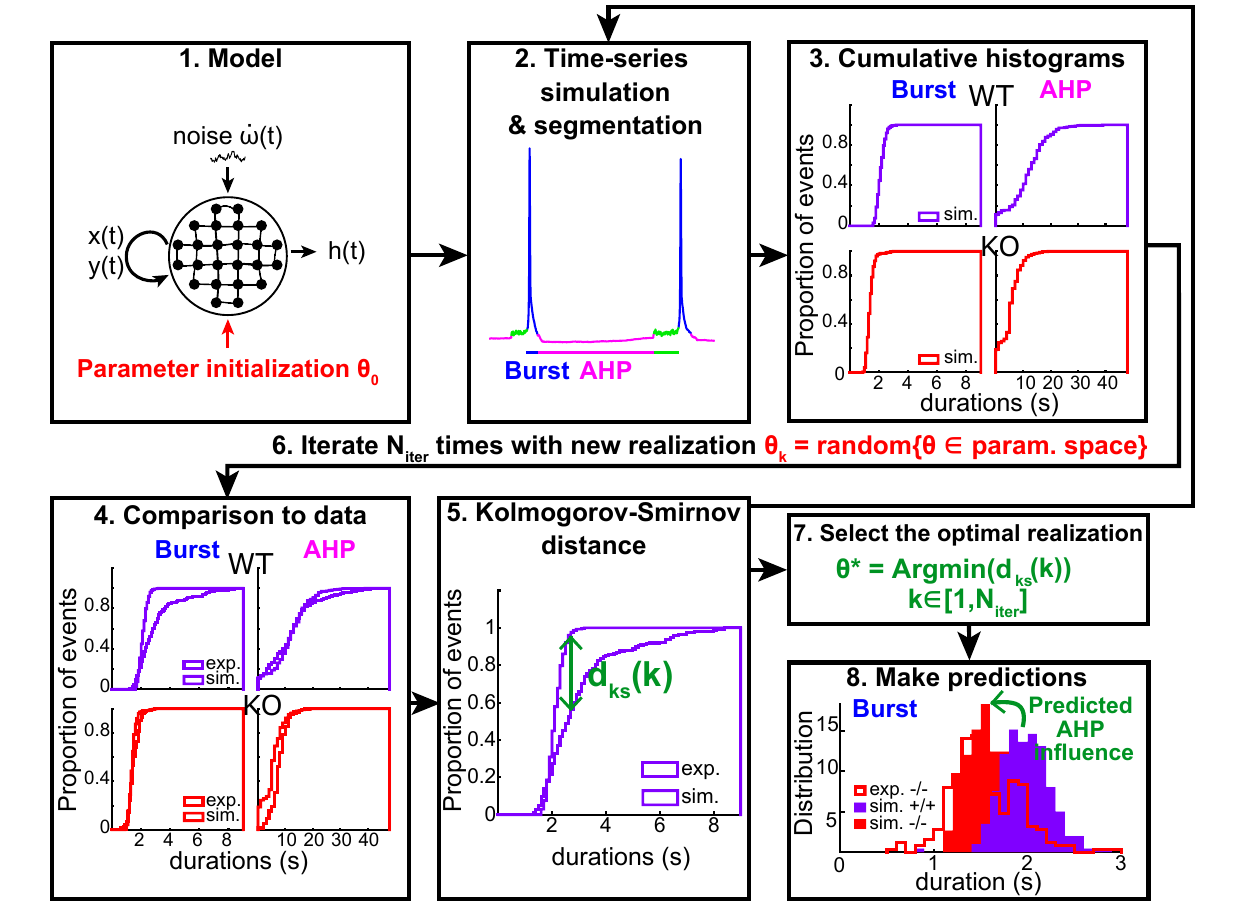}
\caption{\textbf{Parameter fitting pipeline.} {\bf 1.} Model with 8 free parameters (vector $\theta$) to be fitted. {\bf 2.} At each step $k$, run and segment (ex. $5000$s) simulations to generate time series with the parameters $\theta_k$. {\bf 3.} Build the cumulative histograms such as bursting and AHP durations from the numerical simulations. {\bf 4-5.} Compare the simulated histograms to the experimental ones by computing the Kolmogorov-Smirnov distance $d_{ks}(k)$ between the experimental and simulated distributions. {\bf 6.} Iterate at step (2) (ex: $K=1000$ times) or  {\bf 7.} Select the optimal realization $k^*$. {\bf 8.} the model with optimal parameters is used to explore the role any parameter of interest.} \label{fitPipeline}
\end{figure}

The models presented in the previous section based on synaptic short-term plasticity are insufficient to account for electrophysiological mechanisms such as after hyperpolarization (AHP), that requires modeling the voltage properties. Voltage modeling has remained difficult and empirical, due to the difficulty of simplifying and adapting the Maxwell equations. Voltage conduction based on channels has been approached by the phenomenological and empirical Hodking-Huxley models and their successors where the voltage equation is mostly driven by capacitance and resistance in the form $CdV/dt=\sum I_{\text{ions}}$, where the sum account for various ionic exchanges across the membrane ionic channels \cite{hille1978}.\\
To account for recent recordings of bursting obtained in slices, where the astrocyte network was disrupted, the phenomenological model \ref{AHP_model} based on synaptic short-term plasticity and AHP was introduced with a total of 14 parameters, as already discussed above. The model accounts for the refractory period generated by the potassium channels, leading to a hyperpolarization phase. In the absence of a systematic approach to test how this model could reproduce the electrophysiological time series, it is possible to use the entire distributions of bursting and interbursting durations. This criteria turns out to be sufficient to identify all parameters. Fitting a distribution is quite constraining as it is a one dimensional curve. For example, many statistical distributions such as Gaussian, Gamma or Theta functions are characterized by two parameters. Thus for a model containing $2N$ parameters, one possibility is to collect $N$ distributions from experimental data.\\
In the present case, the model \eqref{AHP_model} had 14 parameters, 6 of which had already been calibrated \cite{Tsodyks1997,Holcman_Tsodyks2006,Barak2007,DaoDuc2015} leaving 8 parameters to identify. Accordingly, 4 distributions should be fitted. We chose the distributions of the bursting and AHP durations in a WT and in KO cases.\\
While each individual distribution could be fitted by two parameters, fitting the four histograms of bursting and AHP durations in WT and KO at the same time gave a much more robust constrain on 8 parameters. We shall now describe this fitting procedure to estimate the parameters summarized in the form of a pipeline (Fig. \ref{fitPipeline}). The pipeline consists of a model equation and four histograms obtained by segmenting the electrophysiological time series of both burst and AHP durations in WT and KO. The steps are as follows:
\begin{enumerate}
  \item 8 unknown parameters are represented in a vector {\bf $\theta$} (bottom half of Table \ref{tableParam}). For the initial condition, select a random choice inside the authorized ranges for each parameter value (Table \ref{tableParam} right column).
  \item At step $k$, new values of the 8 parameters in $\theta_k$ are chosen randomly in their authorized range $[A,B]$ (Table \ref{tableParam} right column). Once the parameters are chosen, we run $5000$ simulations from the model \eqref{AHP_model}, leading to time series that are then segmented using the procedure described in subsection \ref{SegmentSim}. The statistics for three epochs burst, AHP and QP durations are then generated.
  \item The segmentation of step 2 allows to build the histograms of burst and AHP durations from the numerical simulations.
  \item To find the optimal parameters, generate histograms $f^{(k)}_{Sim}(t)$ from simulations that will be compared to the experimental ones $f_{Exp}(t)$. The method consist in computing the error associated with the Kolmogorov-Smirnov distance
  \beq
  \epsilon_k = d_{ks} (f_{Exp}(t),f^{(k)}_{Sim}(t)).
  \eeq
With four histograms to fit burst and AHP for the WT and KO datasets, it is possible to average over the four $d_{ks}$ distances. The empirical two-sample Kolmogorv-Smirnov distance is given by the maximum of the difference between the cumulative distribution functions $F_{1,2}$ of the two empirical distributions
\beq
d_{ks}(F_1,F_2) = sup_{x}|F_1(x)-F_2(x)|.
\eeq
\item When the error $\epsilon_k$ is higher than a threshold $T_{fit}$,  then the  iterate restarts at step 2, otherwise the fitting algorithm terminates. To ensure that the algorithm had converged, another solution would be to fix the number of iterations at step 2 (e.g $N_{iter} = 1000$) and to pick the best of all realizations:
\beq
k^*= Argmin_{k \in [1,N_{iter}]}(\epsilon_k),
\eeq
in that case,
\beq
\theta^*= \theta_{k^*}
\eeq
\end{enumerate}
For step 2, this approach was used in the absence of a deterministic strategy to explore the parameter space. Other strategy could be a gradient descent if an energy function was known (Fig. \ref{optimFig}). This procedure is quite generic and could be adapted to other models and any number parameters. The pipeline implementation in Matlab can be found in  \href{www.bionewmtrics.org}{\url{bionewmetrics.pipeline.parameter.calibration}}. The model used in this code is the model \eqref{AHP_model} but it can be replaced by any other model implemented as a Matlab function returning time-series with large amplitude activation events such as bursts.
\begin{center}
\begin{tabular}{l l l l }
& Fixed parameters & Values &  \\
\hline
$\tau$ & Fast time constant for $h$ & 0.05s \cite{Holcman_Tsodyks2006} & \\
$K$ & Facilitation rate & 0.037Hz &(\textit{modified: 0.04Hz in} \cite{DaoDuc2015})\\
$L$ & Depression rate & 0.028Hz &(\textit{modified: 0.037Hz in} \cite{DaoDuc2015})\\
$\tau_r$ & Depression time rate & 2.9s &(\textit{modified: 2-20s in} \cite{DaoDuc2015})\\
$\tau_f$ & Facilitation time rate & 0.9s &(\textit{modified: 1.3s in} \cite{DaoDuc2015})\\
$T$ & Depolarization parameter & 0 &\\
$H_{AHP}$ & End of recovery phase threshold on $h$ & -7.5 &\\
\hline \hline
& Parameters to fit & Fitted values & Authorized range\\
\hline
$\tau_{mAHP}$ & Medium time constants for $h$ & 0.15s & $[0.05,1]$s \\
$\tau_{sAHP}$ & Slow time constants for $h$ & 5s & $[1,20]$s\\
$J$ & Synaptic connectivity & 4.21 & $[3,5]$ \cite{Barak2007}\\
$X$ & Facilitation resting value & 0.08825 & $[0,0.2]$\\
$\sigma$ & Noise amplitude & 3 & $[0.1,10]$\\
$T_{AHP}$ & Undershoot threshold & -30 & $[-40,-5]$\\
$Y_{AHP}$ & Recovery phase threshold & 0.85 & $[0.75,0.95]$\\
$Y_{h}$ & end of AHP phase threshold & 0.5 & $[0.45,0.55]$\\
\hline
\end{tabular}
\end{center}\captionof{table}{Model \eqref{AHP_model} parameters}\label{tableParam}
\subsection{Possible improvements of the fitting pipeline}
There are several possible improvements for the fitting pipeline: first, it will be soon possible to use fully automatized softwares as described in subsection \ref{pipelinemeth}. Second, the fit quality between the data and the simulation depends on the chosen metric. Here we used the Kolmogorov-Smirnov distance but there are other distances that could be used to better capture similarities and differences between the histograms. One possibility would be to use the Wassertein distance \cite{ruschendorf1985wasserstein,vallender1974calculation}, which is defined in optimal transport theory \cite{villani2014OT,villani2008optimal}.  This distance is also called the earth mover's distance because can be seen as the minimal workload required to move one pile of soil to another one with specific shapes. The piles of soil being metaphors for abstract distributions, allowing to compar histograms. Since this method is already used in various theoretical fields, several implemented versions of this metric are openly available online (see for example \cite{wassersteinDistImplementation} for a Matlab implementation, the scipy library in Python also provides an implementation of this distance, see \href{https://docs.scipy.org/doc/scipy/reference/generated/scipy.stats.wasserstein_distance.html}{\url{scipy.stats.wasserstein_distance}}).\\
Finally, it is often difficult to find an optimal strategy for the exploration of the parameter space especially in high dimension. A random exploration is greedy and not efficient. When the space is the produce of intervals, the  unit space parameter is a cube in high dimension. One possibility would be to use a uniform point distribution with a resolution $r$. Then refining the exploration around the minimum point is possible by a subdivision with a resolution $r/2$. This could lead to an accelerated search of the optimal parameters. It remains an open question in general to know whether there is a unique solution for such complex dynamical systems. The optimal solution could also be located in a shallow minimum where neighboring points provide as well good results.
\subsection{Network models to decipher the role of each parameter}
After the calibration step, network model such as the one presented by equations \eqref{AHP_model} can be used to decipher the respective influence of AHP, synaptic noise or membrane depolarization on neuronal bursting activity, that are all affected by the disruption of the astrocyte network. Specifically, this step can be performed by running simulations with all the parameters obtained in the case of the WT fit and then changed to the KO values only the parameters corresponding to the condition to be explored. \\
The distribution of the simulated burst and IBI durations can be compared to ones of the the KO data. Using this approach, it turns our that AHP was sufficient to explain the change in burst and IBI durations observed in the KO data, while the increase of synaptic noise had some effect on burst durations but not much on IBI and the membrane depolarization had no effect on either duration. To conclude, these results clarify the physiological mechanisms underlying the regulation of neuronal bursting by astrocytes in a way that would not have been possible experimentally.
\section{Discussion and perspectives}
We illustrated in this book chapter how data analysis can benefit from neuronal network modeling based on synaptic short-term plasticity. We presented a generic signal processing approach and the associated algorithms to segment large amplitude bursting events from a background, characterized by a smaller amplitude. Indeed, this approach requires that burst-like events should be well separated and differ sufficiently from the background signal. Although this segmentation algorithm is quite elementary, when the baseline can vary or when multiple bursting event can co-exist, several adjustment are necessary, as we saw for the case of calcium dynamics in astrocytes.\\
We did not review here amplitude events that need to be segmented and replaced in real time: this could be the case for artifacts located in the EEG channels \cite{chavez2018surrogate,dora2022adaptive}. For example, during general anesthesia, large artifacts can be generated due to eye motion (EOG), muscle contraction (EMG) or simply because electrodes are transiently moved. Removing these artifacts require not only to segment the exact epoch, but also to replace them by a more physiological signal. The method presented in this chapter is insufficient for this task and other approaches based on wavelet transform and coefficient transportation have been developed for real-time brain monitoring \cite{dora2022adaptive}.\\
A new generation of segmentation problems involve time-frequency dynamics, relevant to detect short high frequencies responses such as ripples \cite{buzsaki_rhythms_2006,weiss2020ripples}. It is also the case for the segmentation of transient epileptoid signs during children anesthesia \cite{rigouzzo_relationship_2008,rigouzzo_eeg_2019}. This approach requires different techniques such as comparison with a pre-existing alphabet of motifs using more elaborated projection methods such as independent component analysis.\\
Finally, computational models have now matured enough so that they play a key role in our understanding, quantification and prediction of neuronal network dynamics, allowing to decipher for example the contribution of short-term synaptic plasticity from other biophysical parameters. They also served to study the emergence of oscillation regimes from non-synchronous oscillators, or to quantify the contribution of membrane fluctuation in rhythm genesis\cite{Guerrier2015,del2018breathing}.\\
Probably, the next generation of models will become avatars of physiological processes. To that end, they should have well calibrated parameters to reproduce the statistics of physiological realizations. Obtaining such calibrations remains difficult for medical applications such as real-time brain prediction during performing tasks, anesthesia, coma or simply to predict an epileptic crisis sufficiently in advance. Another difficulty to move from neuronal network representation to models involving brain regions is that this complexity remains difficult to handle due to large amount of parameters, but once such calibrations would be possible by comparing distribution, we can foresee real-time applications and progress in managing or controlling the brain under various conditions.
\section*{Acknowledgments}
DH research is supported by a grant ANR NEUC-0001, a Memolife grant and the European Research Council (ERC) under the European Union’s Horizon 2020 research and innovation programme (grant agreement No 882673). European Research Council (Consolidator grant 683154) (NR) European Union’s Horizon 2020 research and innovation program (Marie Sklodowska-Curie Innovative Training Networks, grant 722053, EU-GliaPhD) (NR) Ligue Francaise contre l’Epilepsie (ED)
Aviesan (ED). French Research Ministry (ED386, Ecole Doctorale de Sciences Mathématiques Paris centre). LZ fellowship was supported Fondation pour la Recherche Medicale (FRM FDT202012010690).
\section*{Competing interests}
The authors declare no competing financial interests.
\bibliographystyle{ieeetr}
\bibliography{bibliolzfinal,biblio2,bibliography,bibElena,zotero}
\end{document}